\documentclass{JHEP3}

\usepackage{graphicx,times}
\usepackage{multirow}
\usepackage{amsmath,amssymb}

\newcommand{\psib}{\overline{\psi}}

\title{Fermion bag approach to the sign problem in \\
strongly coupled lattice QED with Wilson fermions}
\author{Shailesh Chandrasekharan and Anyi Li\\
  Department of Physics, Box 90305, Duke University,
  Durham, North Carolina 27708, USA
  E-mail: \email{sch@phy.duke.edu,anyili@phy.duke.edu}}

\abstract{We explore the sign problem in strongly coupled lattice QED with one flavor of Wilson fermions in four dimensions using the fermion bag formulation. We construct rules to compute the weight of a fermion bag and show that even though the fermions are confined into bosons, fermion bags with negative weights do exist. By classifying fermion bags as either simple or complex, we find numerical evidence that complex bags with positive and negative weights come with almost equal probabilities and this leads to a severe sign problem. On the other hand simple bags mostly have a positive weight. Since the complex bags almost cancel each other, we suggest that eliminating them from the partition function may be a good approximation. This modified partition function suffers only from a mild sign problem. We also find a simpler model which does not suffer from any sign problem and may still be a good approximation at small and intermediate values of the hopping parameter. We also prove that when the hopping parameter is strictly infinite all fermion bags are non-negative.}


\begin{document}

\section{Introduction}

Strongly correlated many body fermion problems is an exciting area of research today \cite{Subedi2008,Campbell2009}. The main theoretical challenge in the field is to compute physical quantities of interest from first principles. Most methods that are currently used involve approximations that can be justified only in some regions of the parameter space. For problems where none of these approximations can be justified, the computational challenge is daunting. The Monte Carlo method is the only method which may be reliable in such cases. Unfortunately, this method also suffers from sign problems that arise due to the quantum nature of the underlying system \cite{Zaanen2008,MihailoCubrovic2009}. The final answer usually depends on delicate cancellations between many different quantum amplitudes which the Monte Carlo approach is unable to accomplish efficiently. The physics of nuclear matter and strongly correlated electronic systems are classic examples where the sign problem has hindered progress. Attempts to circumvent or solve the sign problem continues to be an important area of research and is also the focus of the current work.

While a general solution to sign problems may not exist \cite{Troyer:2004ge}, solutions have been found in specific cases when problems are reformulated using new variables. For example, while bosonic quantum field theories with a non-zero chemical potential suffer from a sign problem in the conventional formulation \cite{aarts:131601}, in the world line approach these sign problems disappear \cite{Endres:2006xu,Chandrasekharan:2008gp}. Even in fermionic quantum field theories, where the origin of the sign problem is the Pauli principle, new solutions are beginning to appear. In the conventional approach fermions are integrated out and the partition function is written in terms of bosonic degrees of freedom with a Boltzmann weight equal to the determinant of a matrix \cite{PhysRevB.34.7911}. If this determinant is non-negative then the sign problem is absent and today such problems can be solved using the popular hybrid Monte algorithm \cite{Duane:1987de} and its variants \cite{Luscher:2010ae}. On the other hand in many interesting cases the determinant can be negative or even complex. In such cases the conventional approach offers little hope for further progress. Recent research has shown that the world line formulations offer an alternative approach. Instead of integrating out the fermions at the beginning, considering their world lines and then re summing over only a limited class of these configurations leads to new solutions of the sign problems \cite{Karsch:1988zx,PhysRevLett.83.3116}. The idea of using the world line aproach in two dimensional lattice field theories which usually do not suffer from sign problems has a long history \cite{Salmhofer:1991cc,Gattringer:2007em,Wolff:2007ip,Wolff:2008xa,PhysRevD.80.071503}. Recently these developments have been unified under the framework called the ``fermion bag'' approach which shows that the new solutions to fermion sign problems can emerge in any dimension \cite{PhysRevD.82.025007}. Basically one identifies independent dynamical regions over which the fermions naturally hop. These dynamical regions, called fermion bags, behave like non-local degrees of freedom. The weight of a fermion bag is just the path integral inside the fermion bag. When field theories are written in terms of fermion bags, sign problems may be absent since the weight of the fermion bags can be non-negative. The fermion bag approach allow us to solve some problems that seemed difficult or impossible in the conventional approach \cite{PhysRevD.82.025007}, thanks to new algorithms \cite{Adams:2003cc}. Being a relatively new idea not many examples have been studied and more work is necessary to understand the potential of the method.

In this work we construct the fermion bag approach to four dimensional lattice QED with one flavor of Wilson fermions at strong gauge couplings. In a sense this is an extension of previous work in two \cite{Salmhofer:1991cc} and three dimensions \cite{PhysRevD.80.071503}. Wilson fermions contain a parameter called the hopping parameter referred to here as $\kappa$. It is well known that the determinant of the one flavor Wilson Dirac operator in the background of a strongly fluctuating gauge field configuration can be negative for some values of $\kappa$. Hence the conventional approach suffers from a sign problem in this region. Recently it was shown that the sign problem is absent in three dimensions when the partition function is written in terms of fermion bags \cite{PhysRevD.80.071503}. Is this true in four dimensions? The current work was motivated by this question. At strong gauge couplings fermions are confined into bosons and fermion bags are regions where these bosons hop around. The weight of the bag is then a sum over all paths the fermions can explore within the bag while remaining confined. Since the fermions are always paired there is a possibility that the bags will have a non-negative weight. However, we show here that this is {\em not} the case. Fermion bags with negative weight do exist, suggesting that the underlying bosonic model remains frustrated.

Although the fermion bag approach does not solve the sign problem, we can learn about the nature of the sign problem and some practical solutions from it. First, we can analytically prove that fermion bags with non-negative weights only contribute at $\kappa = \infty$. Thus, the fermion bag approach is able to solve the sign problem at this special point, while the conventional approach has a very severe sign problem there. Second, we find that at small $\kappa$ most bags that contribute have a positive weight. Negative weight bags begin to enter the partition function only for $\kappa > \kappa_c$ as in the conventional approach. Third, we find that large bags which are topologically simple (to be explained later) are also almost always positive. Large complex bags on the other hand have both positive and negative weights with almost equal probability. This creates a severe sign problem if they are allowed in the partition function. However, to a good approximation they seem to cancel each other and the partition function function is dominated only by simple bags. If one assumes this reasoning to be correct one obtains a new model that seems to capture at least some the interesting physics of the original model. This method of identifying new models by focusing on a class of fermion bags which capture important physics while being practically solvable may turn out to be one of the main advantages of the fermion bag approach.

Our paper is organized as follows. In section \ref{signdet} we briefly review the sign problem in strongly coupled lattice QED with one flavor of Wilson fermions in the conventional approach. In section \ref{fbagrules} we develop the fermion bag approach and construct diagrammatic rules to compute the weight of a fermion bag. In section \ref{signfbag} we classify bags as simple and complex and compute the weights of some small bags. We give examples of bags with negative weights. We also find the distribution of simple and some complex bags and use it to justify that complex bags do not contribute to the partition function. In section \ref{posfbags} we contruct a model without a sign problem that most likely contains the physics of parity breaking. We also give an analytic proof that the weight of fermion bags at $\kappa = \infty$ are non-negative. Section \ref{conc} contains our conclusions.

\section{Sign Problem in the Determinant Approach}
\label{signdet}

Let us briefly review the sign problem in the conventional approach to strongly coupled lattice QED with one flavor of Wilson fermions. The partition function is given by
\begin{equation}
Z = \int [d\psib \ d\psi] [d\phi] \exp(-S[\psib,\psi,\phi])
\end{equation}
where the Wilson fermion action is given by
\begin{equation}
S = - \sum_{x,\alpha}\ 
\Big(\psib_x \Gamma^\alpha_+ \mathrm{e}^{i\phi_{x,\alpha}}\psi_{x+\alpha} 
\ +\  
\psib_{x+\alpha} \Gamma^\alpha_-  \mathrm{e}^{-i\phi_{x,\alpha}}\psi_{x}\Big) 
\ +\  \frac{1}{k} \sum_x \psib_x\psi_x
\end{equation}
with the definition $\Gamma^\alpha_\pm = (1 \pm \gamma_\alpha)/2$. We denote the four Hermitian Dirac matrices as $\gamma_\alpha, \alpha = 1,2,3,4$. We also define $\gamma_5 = -\gamma_1\gamma_2\gamma_3\gamma_4$ for later convenience. For explicit calculations we will use the chiral representation in which
\begin{equation}
\gamma_\alpha = \left(\begin{array}{cc} 0 & \tau_\alpha \cr
\tau_\alpha^\dagger & 0 \end{array}\right),\ \ 
\gamma_5 = \left(\begin{array}{cc} I & 0 \cr
0 & -I \end{array}\right).
\end{equation}
The four  $2\times 2$ matrices $\tau_\alpha$ are defined by $(i\vec{\sigma},I)$ in the four vector notation. Note that $\vec{\sigma}$ are the three Pauli matrices. The lattice fields $\psi_x$ and $\psib_x$ represent the two independent Grassmann valued four component Dirac spinors on each hyper-cubic lattice site $x$ and $\phi_{x,\alpha}$ is the compact $U(1)$ lattice gauge field. In this work we choose open boundary conditions for convenience. Further note that our definition of $\kappa$ is two times the conventional definition of $\kappa$ \cite{PhysRevD.25.1130}.

The conventional approach is to integrate out the fermions and express the partition function as simply an integral over gauge fields. In this approach the Boltzmann weight of each gauge field configuration is simply the fermion determinant of the Wilson Dirac operator $D_W[\phi]$ in the background of that gauge field. More explicitly.
\begin{equation}
Z = \int [d\phi]\ \mathrm{Det}\Big(D_W[\phi]\Big).
\label{detZ}
\end{equation}
where 
\begin{equation}
(D_W[\phi])_{x,y} = -\sum_\alpha\ 
\delta_{x+\alpha,y}\Gamma^\alpha_+ \mathrm{e}^{i\phi_{x,\alpha}}
\ +\  
\delta_{x,y+\alpha} \Gamma^\alpha_-  \mathrm{e}^{-i\phi_{y,\alpha}}
\ +\  \frac{1}{k} \delta_{x,y}
\end{equation}
The Wilson Dirac operator satisfies the relation $D_W^\dagger \gamma_5 = \gamma_5 D_w $ which can be used to show that eigenvalues of $D_w$ are either real or come in complex conjugate pairs. For $\kappa < 0.25$ all real eigenvalues can be shown to be positive. However, for larger values of kappa there can in principle be an odd number of negative eigenvalues. Hence the determinant can be negative. The negative determinant is necessary to violate the Vafa-Witten theorem \cite{Vafa:1983tf} and allow for the spontaneously breaking of the parity symmetry that is expected to occur for $\kappa > \kappa_c$ \cite{Aoki:1983qi}. One expects $\kappa_c \sim 0.5$ at strong couplings. 

\FIGURE[h]{
\includegraphics[width=0.6\textwidth]{sign_det}
\caption{Average value of the sign of $\mathrm{Det}(D_w)$ on $4^4$ and $6^4$ lattices as a function of $\kappa$ obtained using $1000$ random gauge field configurations. The mild sign problem on $4^4$ lattice for $0.5 > \kappa > 1.4$ is just a finite size effect. \label{fig:sign_det}}
}

In order to show the sign problem we compute the sign of the determinant of $D_W$ on $4^4$ and $6^4$ lattices in the background of a $1000$ random $U(1)$ gauge field.  We plot the average value of this sign as a function of $\kappa$ in figure~\ref{fig:sign_det}. As expected the determinant approach encounters a severe sign problem when $\kappa > \kappa_c \sim0.5$. The sign problem continues to be severe even at $\kappa = \infty$. In this work we construct the fermion bag approach to this problem.

\section{Fermion Bag Approach}
\label{fbagrules}

At strong coupling we can first perform the link integral over the gauge field connecting $x$ and $x+\alpha$ exactly to obtain an expansion of the partition function in terms of powers of Grassmann variables on each bond. We get
\begin{equation}
\int \frac{d\phi}{2\pi} \exp(
\psib_x \Gamma^\alpha_+ \mathrm{e}^{i\phi}\psi_{x+\alpha} 
\ +\  
\psib_{x+\alpha} \Gamma^\alpha_-\mathrm{e}^{-i\phi}\psi_{x})
= \sum_{k=0}^4\frac{(\psib_x \Gamma^\alpha_+\psi_{x+\alpha}
\psib_{x+\alpha} \Gamma^\alpha_-\psi_{x})^k}{(k!)^2}.
\label{link}
\end{equation}
We can also expand the exponential of the mass term on each site in terms of powers of Grassmann variables
\begin{equation}
\mathrm{e}^{-\psib\psi/\kappa} = \sum_{n=0}^4 \Big(\frac{1}{\kappa}\Big)^n
\frac{[- \psib_x \psi_x]^n}{n!}.
\end{equation}
Collecting all the Grassmann variables on each site and performing the integration over the Grassmann variables using the identity
\begin{equation}
\int [d\psi][d\psib] 
(\psib)_{i_1}\psi_{j_1} (\psib)_{i_2}\psi_{j_2} (\psib)_{i_3}\psi_{j_3} (\psib)_{i_4}\psi_{j_4} = \varepsilon_{i_1 i_2 i_3 i_4} \varepsilon_{j_1 j_2 j_3 j_4}
\end{equation}
we can rewrite the partition function as a sum over bond variables $k_{x,\alpha}=0,1,2,3,4$ and site variables $n_x=0,1,2,3,4$. We will refer to $n_x$ as the number of monomers on the site $x$ and $k_{x,\alpha}$ as the number of dimers on the bond connecting the site $x$ and $x+\hat{\alpha}$. Thus, in the monomer, dimer representation the partition function given by
\begin{equation}
Z = \sum_{[n,k]} \ \prod_B(\omega_B[n,k])
\end{equation}
where $\omega_B$ is the weight of a ``fermion bag'' $B$ which is simply the set of sites connected by $k_{x,\alpha} \neq 0$. Note that every site belongs to a unique bag and fermions only hop within the sites of the bag. The Boltzmann weight of a fermion bag, $\omega_B$, is the sum over a well defined set of fermion hoppings within the bag. Of course there is no need for $\omega_B$ to be positive. However, in this work we prove that $\omega_B$ is indeed positive when $\kappa = 0$ and $|\kappa|=\infty$. We also find evidence that to a good approximation certain class of fermion bags, which almost always have positive weights, dominate the partition function for a range of values of $\kappa$.

Let us now construct the rules for calculating $\omega_B$. For this purpose we define
\begin{equation}
S_{+,\alpha} =  \frac{1}{\sqrt{2}}\left(\begin{array}{cc} I & +\tau_\alpha \cr
0 & 0 \end{array}\right)
\ \ \ \ 
S_{-,\alpha} =  \frac{1}{\sqrt{2}}\left(\begin{array}{cc} I & -\tau_\alpha \cr
0 & 0 \end{array}\right)
\end{equation}
it is easy to show that $\Gamma^\alpha_+ = S^\dagger_{+,\alpha} S_{+,\alpha},\ \Gamma^\alpha_- = S^\dagger_{-,\alpha} S_{-,\alpha} $ for every $\alpha$. Further
\begin{equation}
S_{-s_1,\alpha_1} S^\dagger_{s_2,\alpha_2} = \frac{1}{2}
\left(\begin{array}{cc} (I - s_1 s_2\ \tau_{\alpha_1}\tau^{\dagger}_{\alpha_2}) 
& 0 \cr
0 & 0 \end{array}\right) 
= 
\left(\begin{array}{cc} 
R^{s_1s_2}_{\alpha_1,\alpha_2}
& 0 \cr
0 & 0 \end{array}\right)
\label{sprod} 
\end{equation}
where $R^{s_1 s_2}_{\alpha_1,\alpha_2} \equiv (I-s_1s_2\tau_{\alpha_1}\tau^\dagger_{\alpha_2})/2$ is a $2\times 2$ matrix which can be parametrized as $F^{s_1s2}_{\alpha_1,\alpha_2} \exp(i\mathbf{n}_{\alpha_1,\alpha_2}^{s_1s_2}\cdot\vec{\sigma}(\pi/4))$. table~\ref{tab1} lists the values of $F$ and $\mathbf{n}$ for various possibilities.

\TABLE[h]{
\begin{tabular}{c|c|c|c}
\hline
\hline
$\alpha_1$ & $\alpha_2$ & $F^{s_1s_2}_{\alpha_1,\alpha_2}$ & $\mathbf{n}^{s_1s_2}_{\alpha_1,\alpha_2}$ \\
\hline
$\alpha$ & $\alpha$ & $\frac{1}{2}(1-s_1s_2)$ & $0$ \\
4 & $i=1,2,3$ & $\frac{1}{\sqrt{2}}$ & $n_k=s_1s_2\delta_{ik}$ \\
$i=1,2,3$ & 4 & $\frac{1}{\sqrt{2}}$ & $n_k=-s_1s_2\delta_{ik}$ \\
$i=1,2,3$ & $\substack{j=1,2,3\\j\neq i}$ & $\frac{1}{\sqrt{2}}$ &
$n_k = -s_1s_2\epsilon_{ijk}$ \\
\hline
\end{tabular}
\caption{\label{tab1} Values of $F^{s_1s_2}_{\alpha_1,\alpha_2}$ and $\mathbf{n}^{s_1s_2}_{\alpha_1,\alpha_2}$ that enter the definition of $R^{s_1 s_2}_{\alpha_1,\alpha_2}$.}
}

Note that $R^+_{\alpha,\alpha} = 0$ and $R^-_{\alpha,\alpha} = I$, while for all other values of $\alpha_1$ and $\alpha_2$, the matrix $R^{s_1s_2}_{\alpha_1\alpha_2}$ is $(1/\sqrt{2})$ times a $(1/2,0)$ representation of an $O(4)$ rotation matrix. Using these relations we can write
\begin{eqnarray}
\psib_x\Gamma^\alpha_+\psi_{x+\alpha}\psib_{x+\alpha} \Gamma^\alpha_-\psi_x
&=& 
i\ \Big[\sum_{k,l}\ (S_{-,\alpha})_{i k}\ (\psi_x)_k (\psib_x)_{l}
\ (S_{+,\alpha}^\dagger)_{lj}\Big]
\nonumber \\
&&
i\ \Big[\sum_{m,n}\ (S_{+,\alpha})_{j m} \ (\psi_{x+\alpha})_{m} \ (\psib_{x+\alpha})_{n}
\ (S_{-,\alpha}^\dagger)_{ni}
\Big]
\nonumber \\
\end{eqnarray}
The integration over the Grassmann variable then leads to specific rules that help to compute the weight $\omega_B$ of a bag. 

The weight of a bag turns out to be the trace of the product of Dirac tensors associated to each site. The explicit form of these Dirac tensors are discussed below. But first is useful to remember the constraint that every site in the bag must satisfy
\begin{equation}
n_x + \sum_\alpha \ k_{x,\alpha} + k_{x,-\alpha} = 4.
\end{equation}
Here we have defined $k_{x, -\alpha} = k_{x-\alpha,\alpha}$. Based on the allowed values of $n_x$, each site in the bag can be one of seven types as shown in table~\ref{tab2}. We call these as type-0,1,2,3,4a,4b and 4c depending on the number of dimers attached to the site. Note that there are three types of sites with four dimers attached to it. We distinguish them because the rules to compute the weights are slightly different for each of them. We also use two types of diagrammatic representation for each vertex: a detailed diagram and a minimal diagram. The detailed diagram shows each fermion line and is helpful in the actual computation, while the minimal diagram just shows the dimers (or a single monomer when no dimers exist on the site). Given the minimal diagram the detailed diagram can be uniquely obtained. 

\TABLE[h]{
\begin{tabular}{c|c|c||c|c|c}
\hline 
Detailed & Minimal & Dirac &
Detailed & Minimal & Dirac 
\\
Diagram & Diagram & Tensor &
Diagram & Diagram & Tensor
\\
\hline 
& & & & & \\
\begin{minipage}[c]{0.15\textwidth}
\begin{center}
\includegraphics[width=0.6\textwidth]{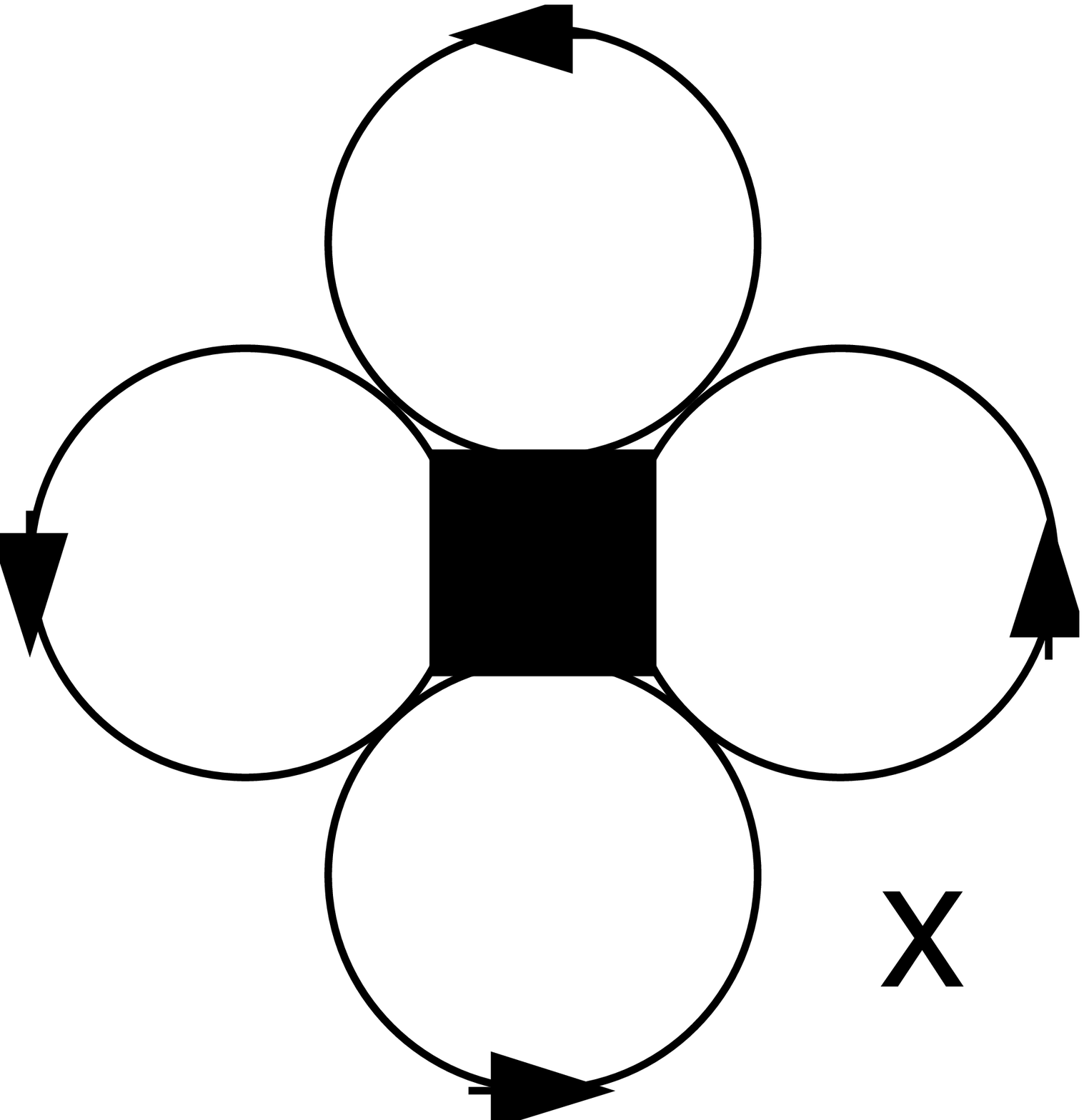}
\end{center}
\end{minipage}
&
\begin{minipage}[c]{0.15\textwidth}
\begin{center}
\includegraphics[width=0.7\textwidth]{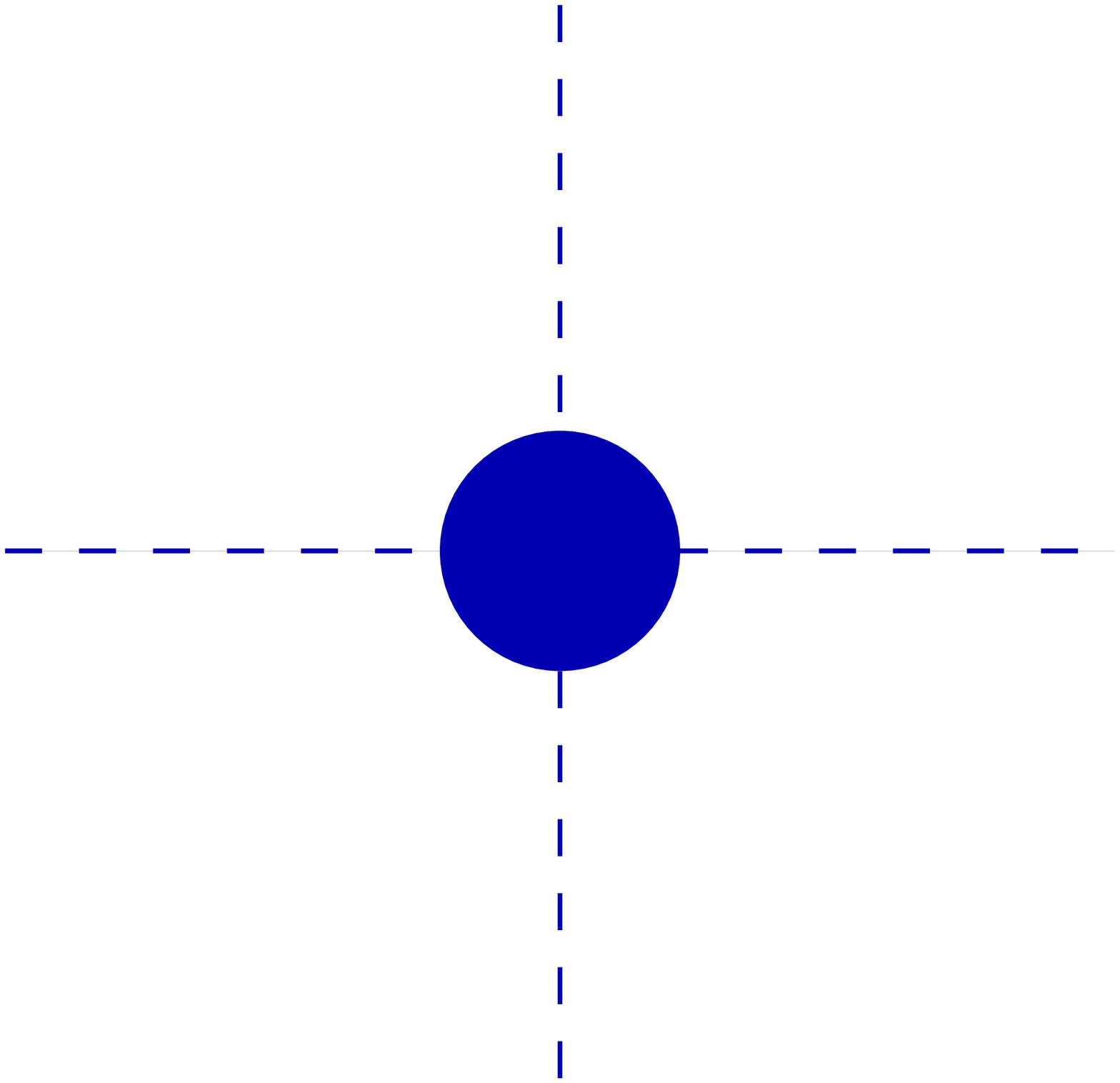}
\end{center}
\end{minipage}
&
$W_0$
&
\begin{minipage}[c]{0.15\textwidth}
\begin{center}
\includegraphics[width=0.7\textwidth]{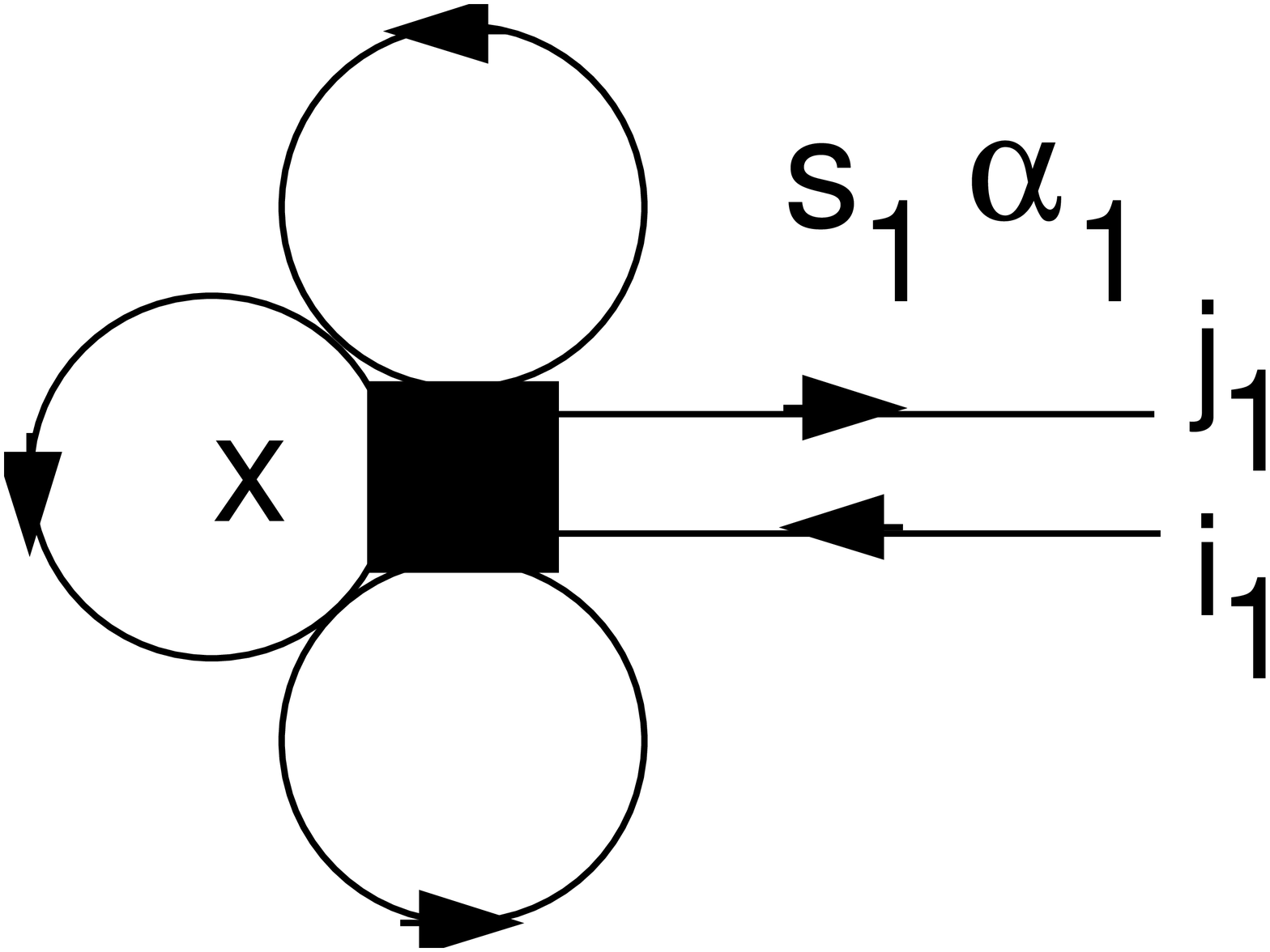}
\end{center}
\end{minipage}
&
\begin{minipage}[c]{0.15\textwidth}
\begin{center}
\includegraphics[width=0.7\textwidth]{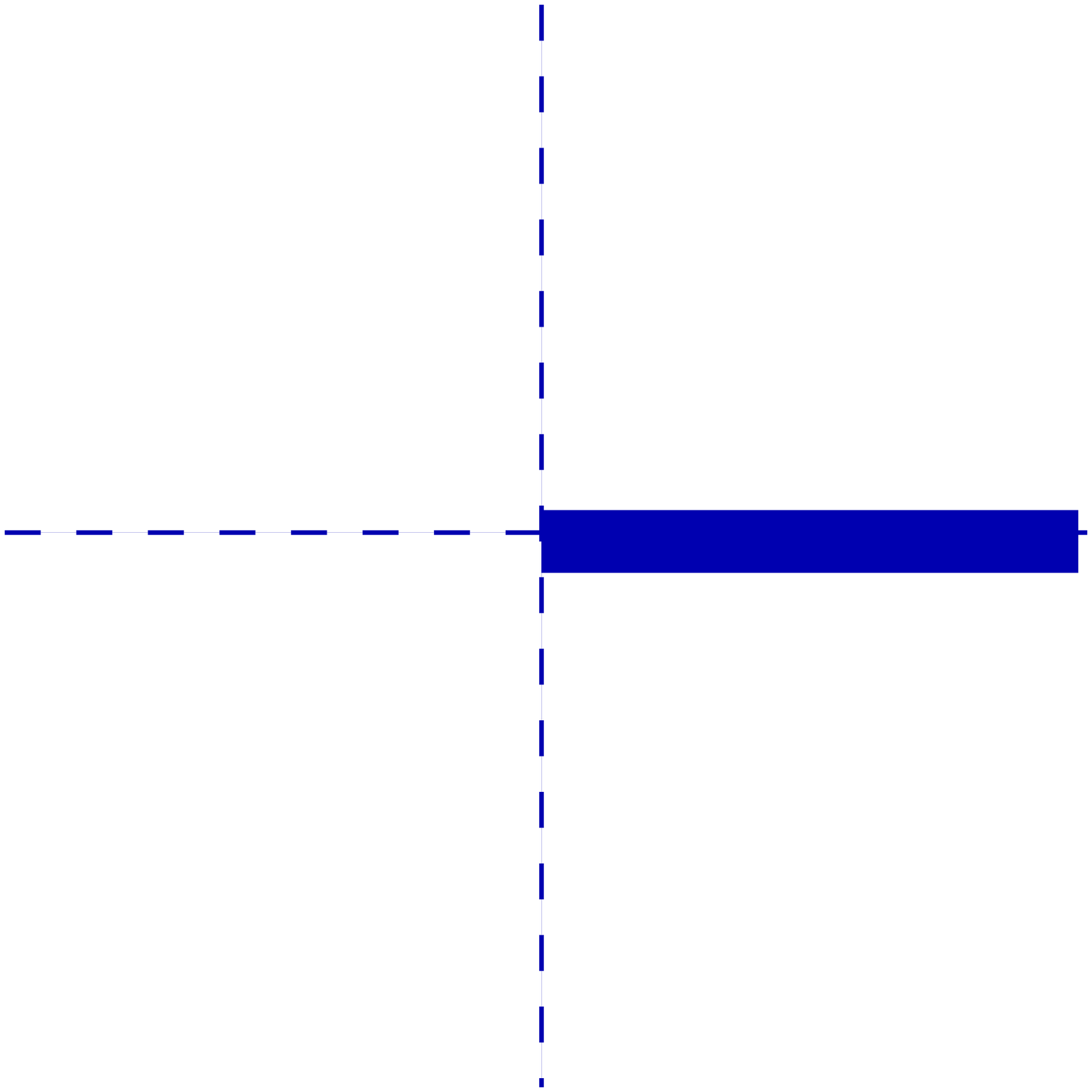}
\end{center}
\end{minipage}
&
$W_1$.
\\
& & & & & \\
\hline
& & & & & \\
\begin{minipage}[c]{0.15\textwidth}
\begin{center}
\includegraphics[width=0.7\textwidth]{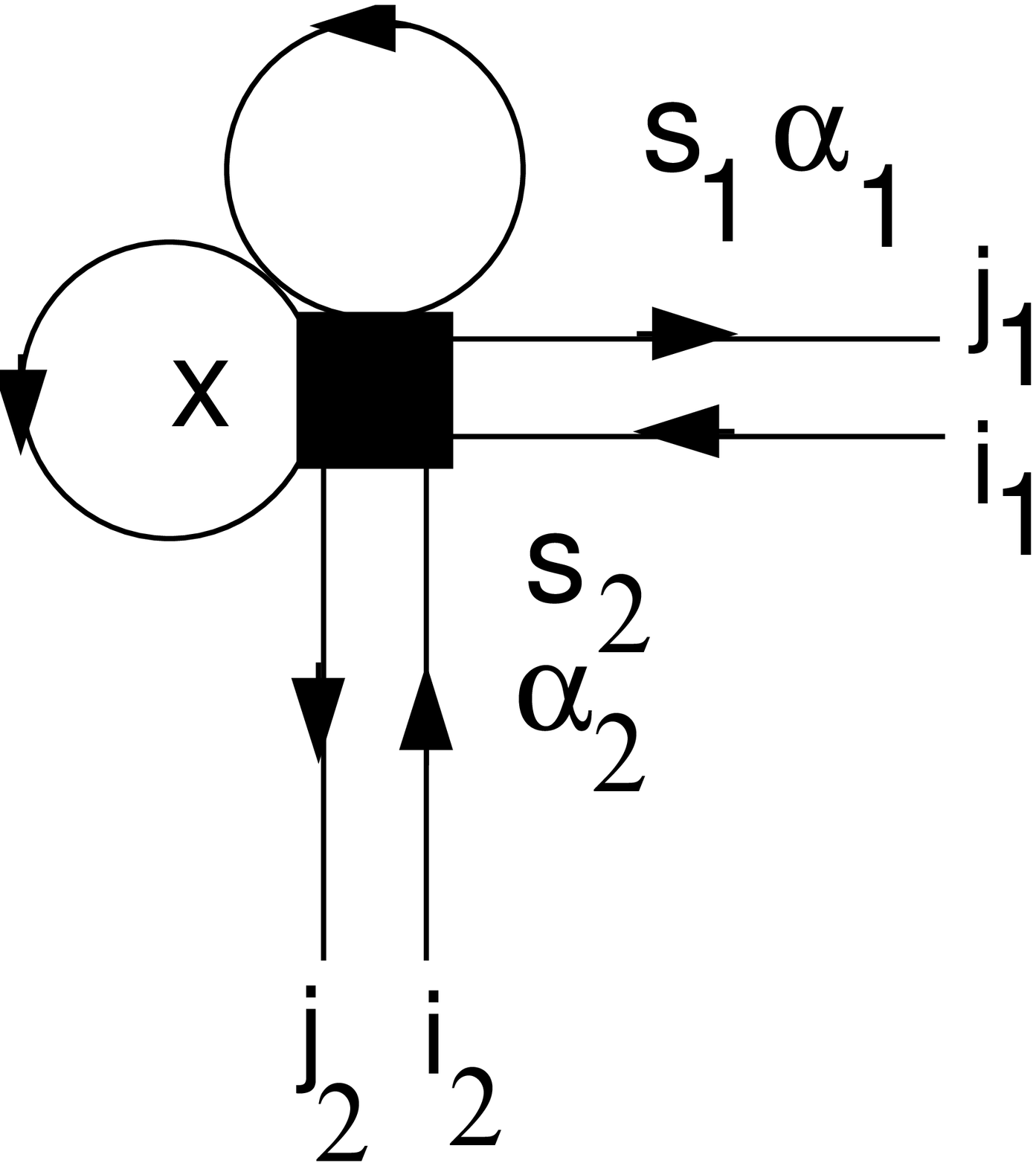}
\end{center}
\end{minipage}
&
\begin{minipage}[c]{0.15\textwidth}
\begin{center}
\includegraphics[width=0.7\textwidth]{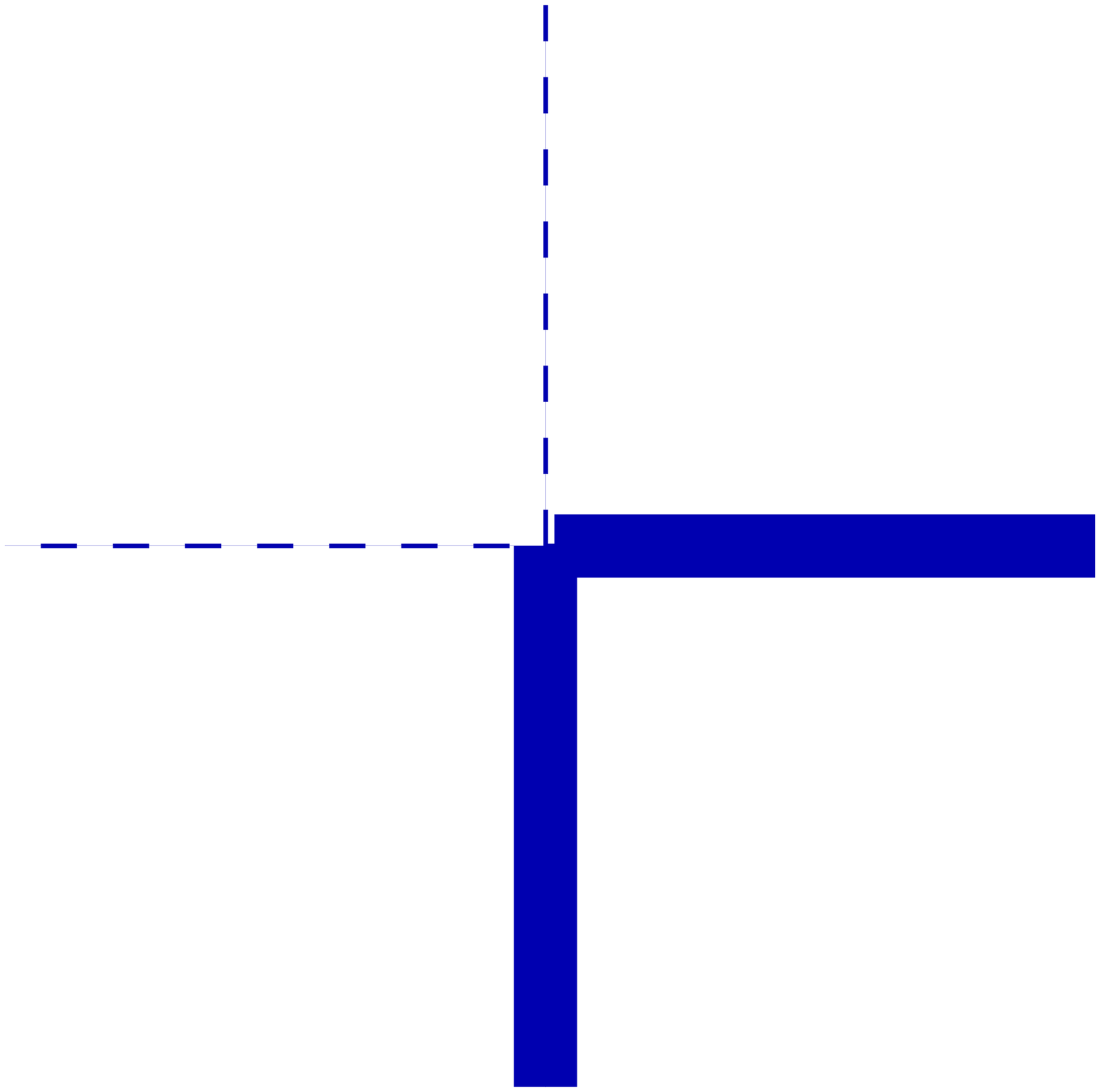}
\end{center}
\end{minipage}
&
$W_2$
&
\begin{minipage}[c]{0.15\textwidth}
\begin{center}
\includegraphics[width=0.7\textwidth]{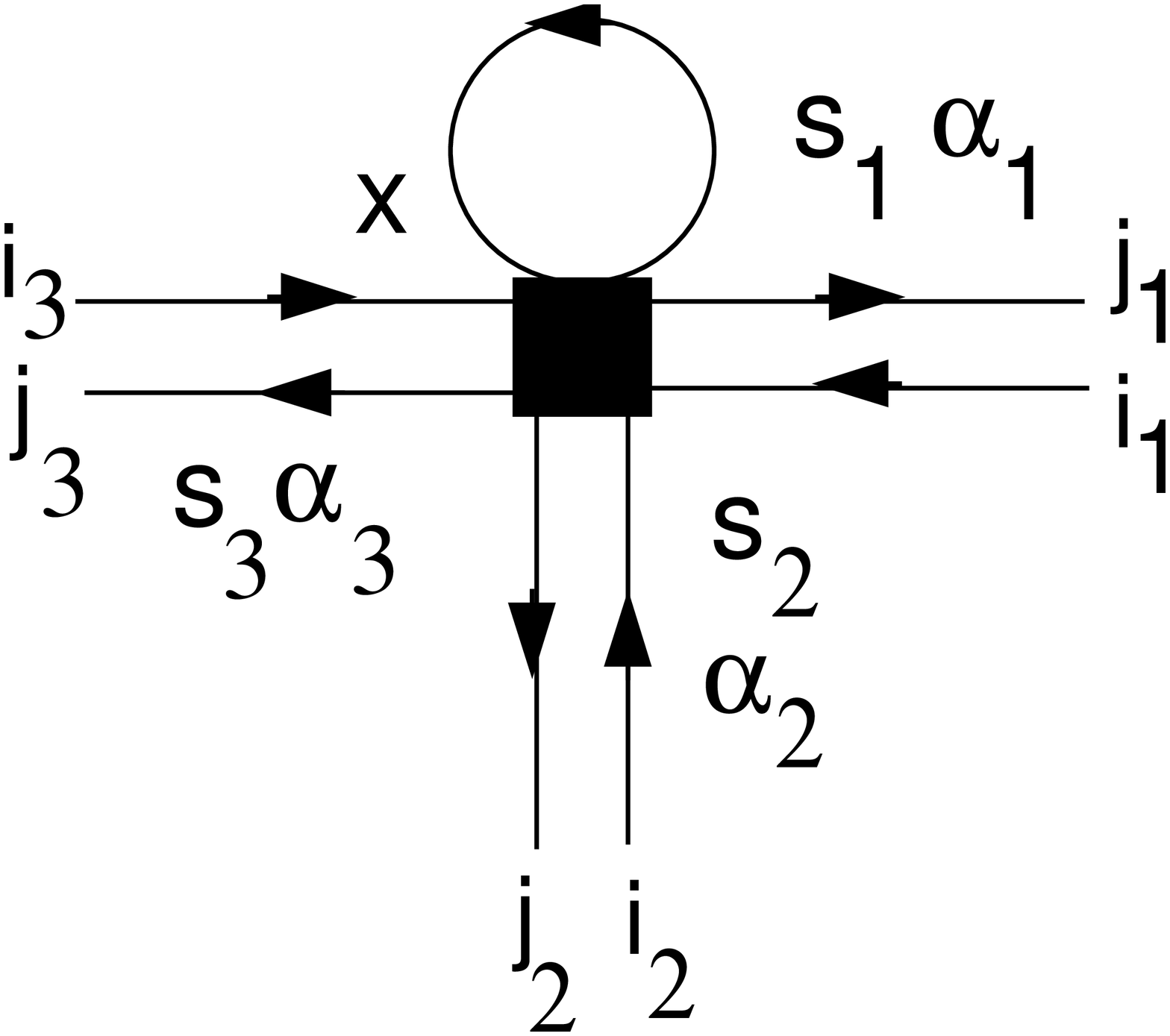}
\end{center}
\end{minipage}
&
\begin{minipage}[c]{0.15\textwidth}
\begin{center}
\includegraphics[width=0.7\textwidth]{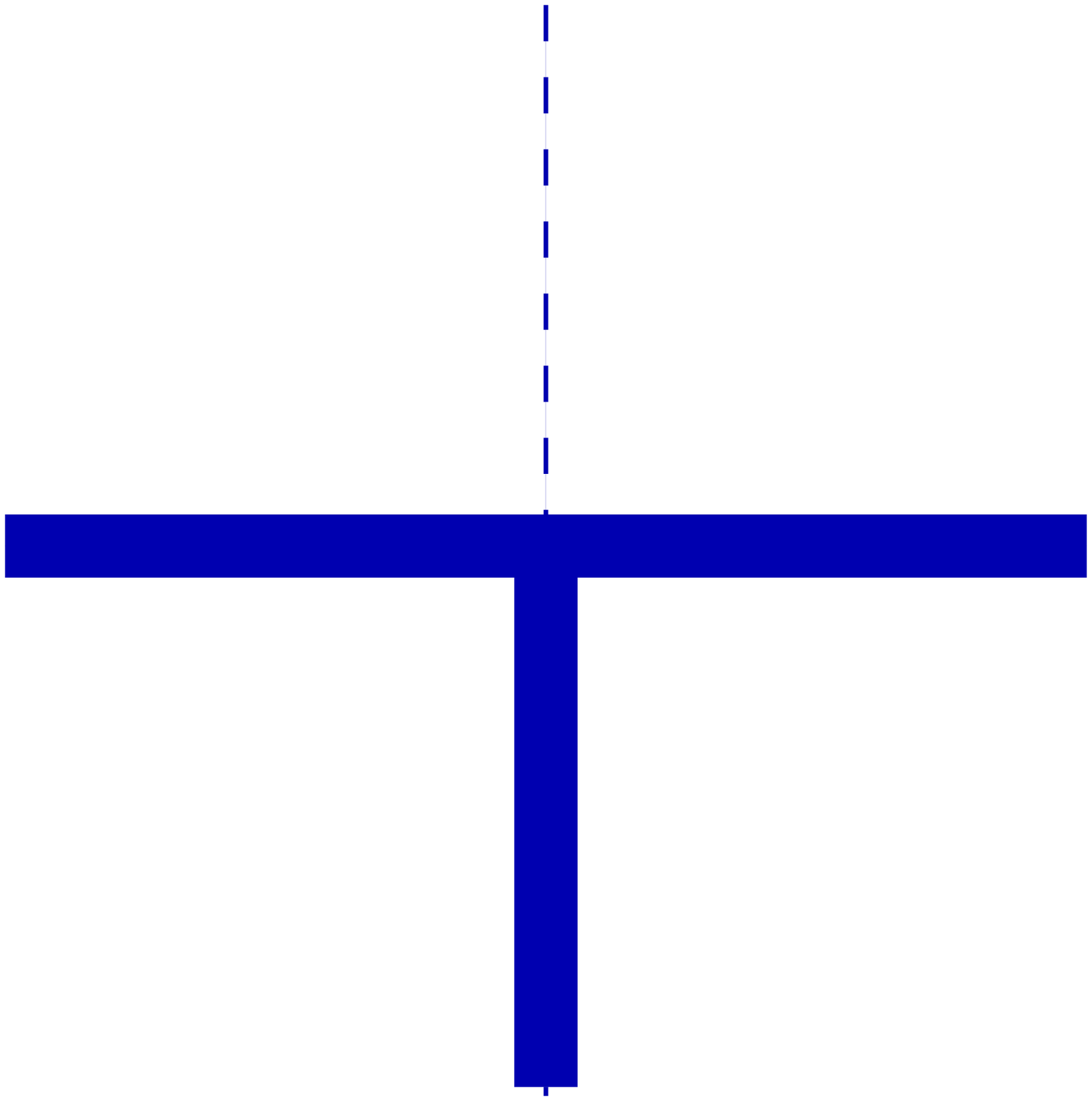}
\end{center}
\end{minipage}
&
$W_3$
\\
& & & & & \\
\hline
& & & & & \\
\begin{minipage}[c]{0.15\textwidth}
\begin{center}
\includegraphics[width=0.75\textwidth]{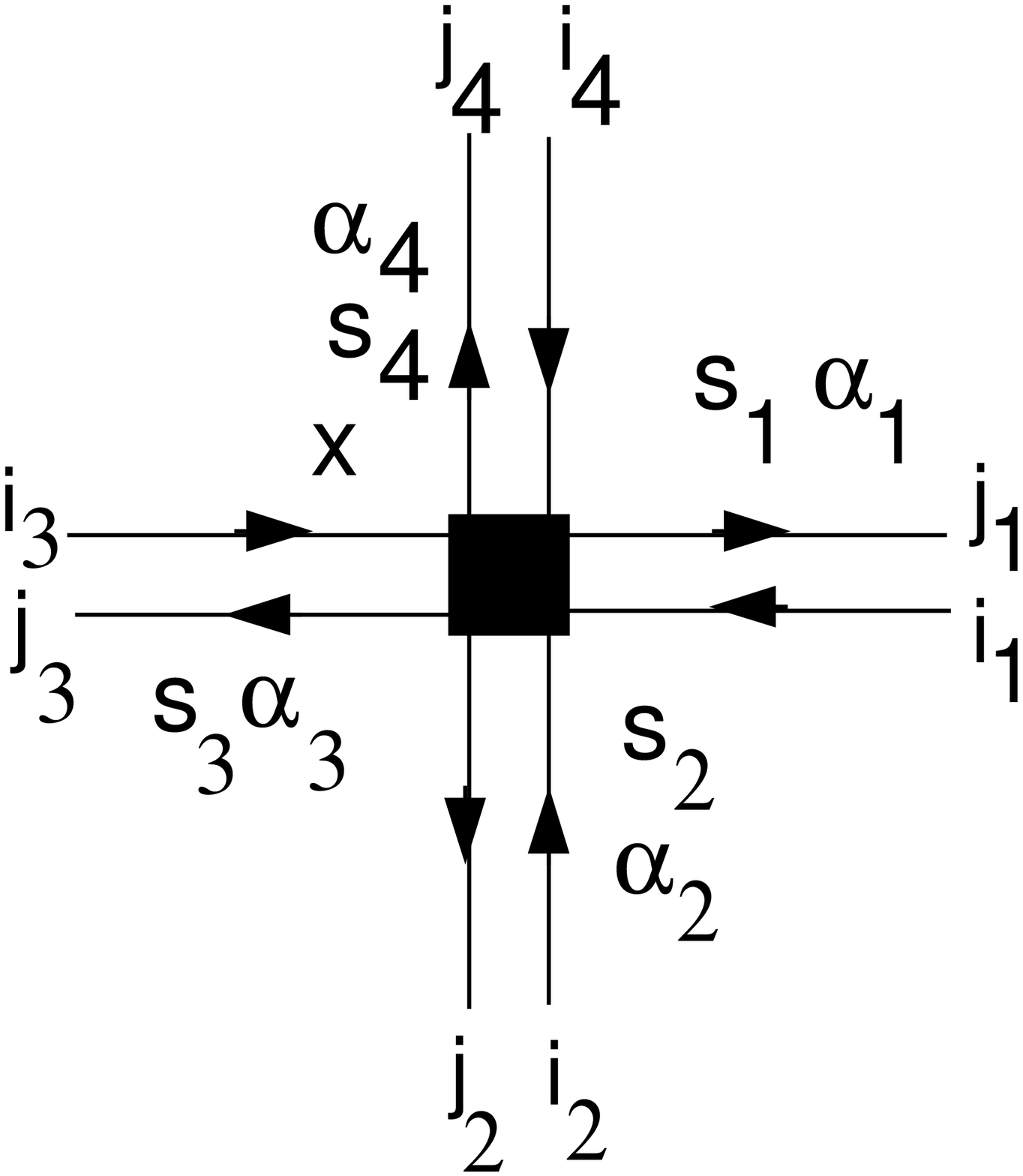}
\end{center}
\end{minipage}
&
\begin{minipage}[c]{0.15\textwidth}
\begin{center}
\includegraphics[width=0.75\textwidth]{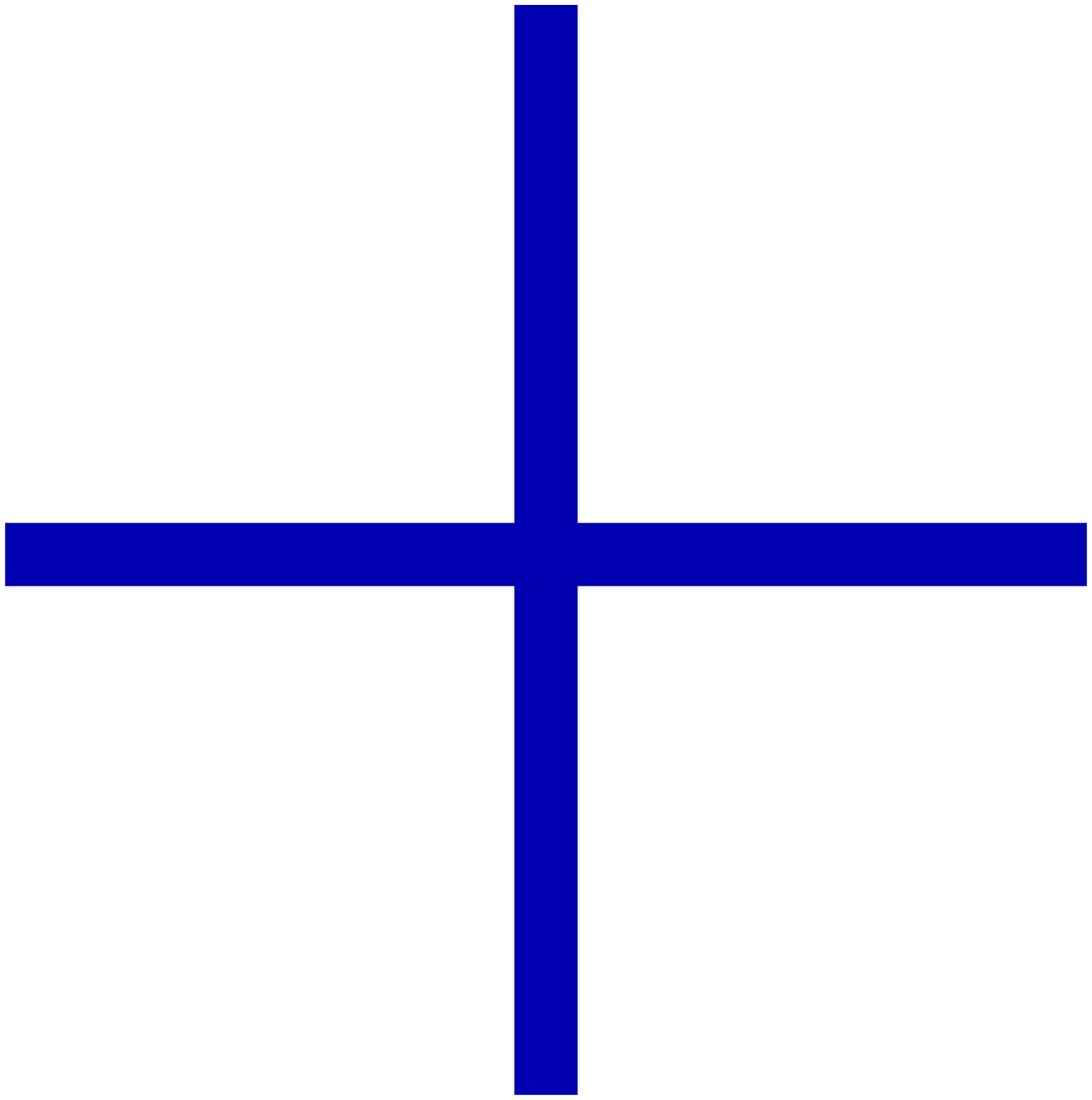}
\end{center}
\end{minipage}
&
$W^a_4$
& 
\begin{minipage}[c]{0.15\textwidth}
\begin{center}
\includegraphics[width=0.75\textwidth]{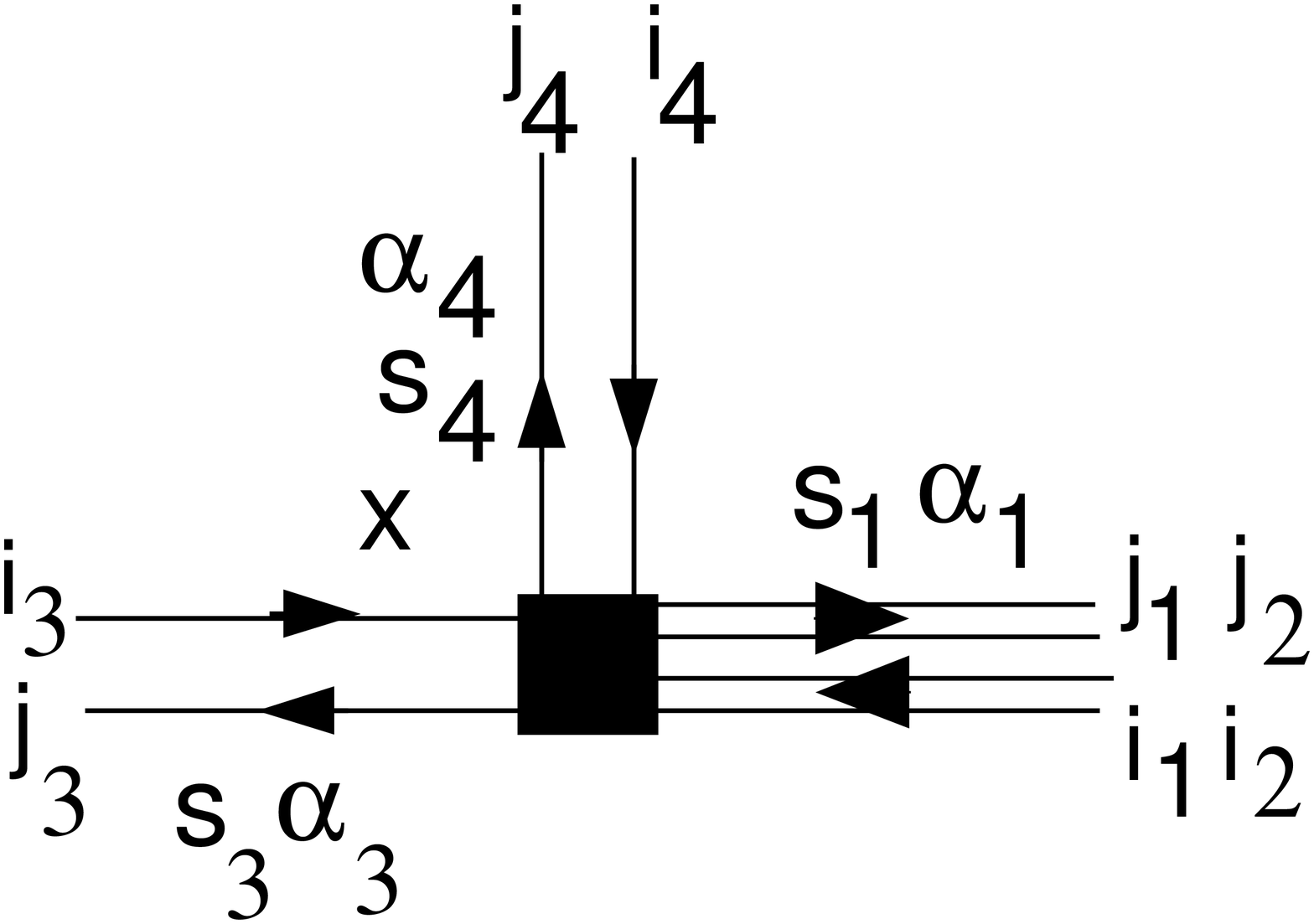}
\end{center}
\end{minipage}
&
\begin{minipage}[c]{0.15\textwidth}
\begin{center}
\includegraphics[width=0.75\textwidth]{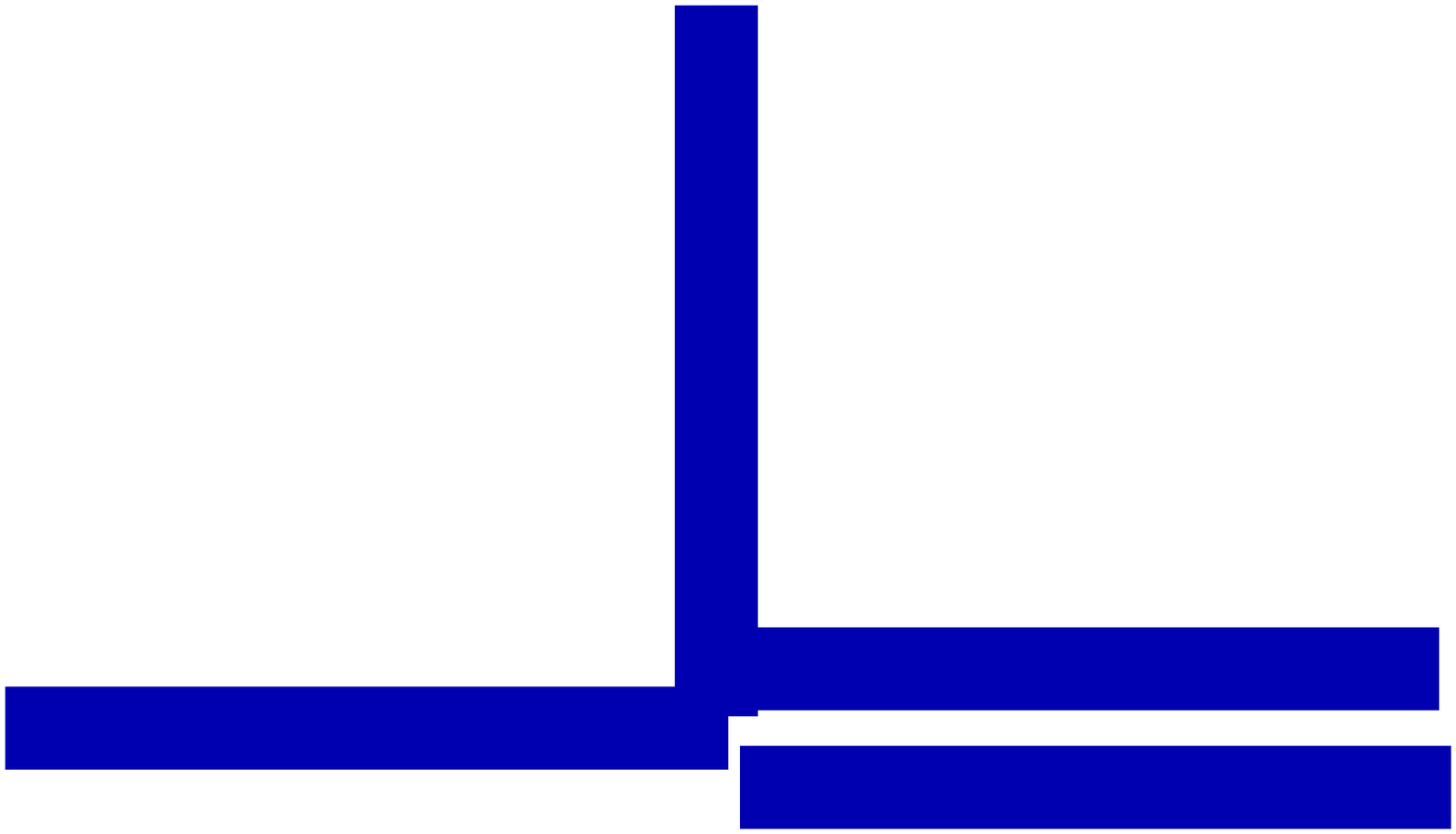}
\end{center}
\end{minipage}
&
$W^b_4$
\\
& & & & & \\
\hline
& & & & & \\
\begin{minipage}[c]{0.15\textwidth}
\begin{center}
\includegraphics[width=0.75\textwidth]{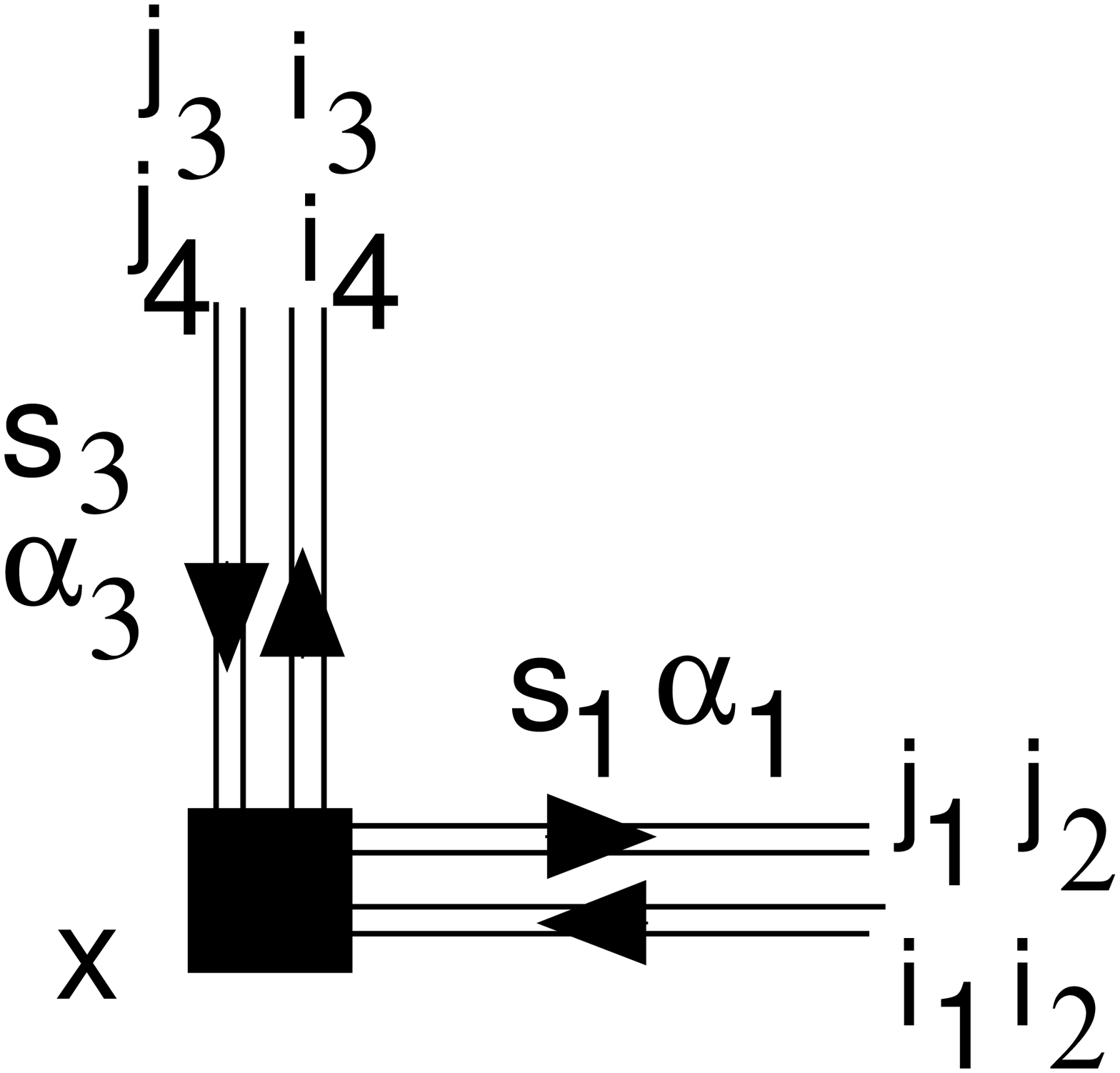}
\end{center}
\end{minipage}
&
\begin{minipage}[c]{0.1\textwidth}
\begin{center}
\includegraphics[width=0.7\textwidth]{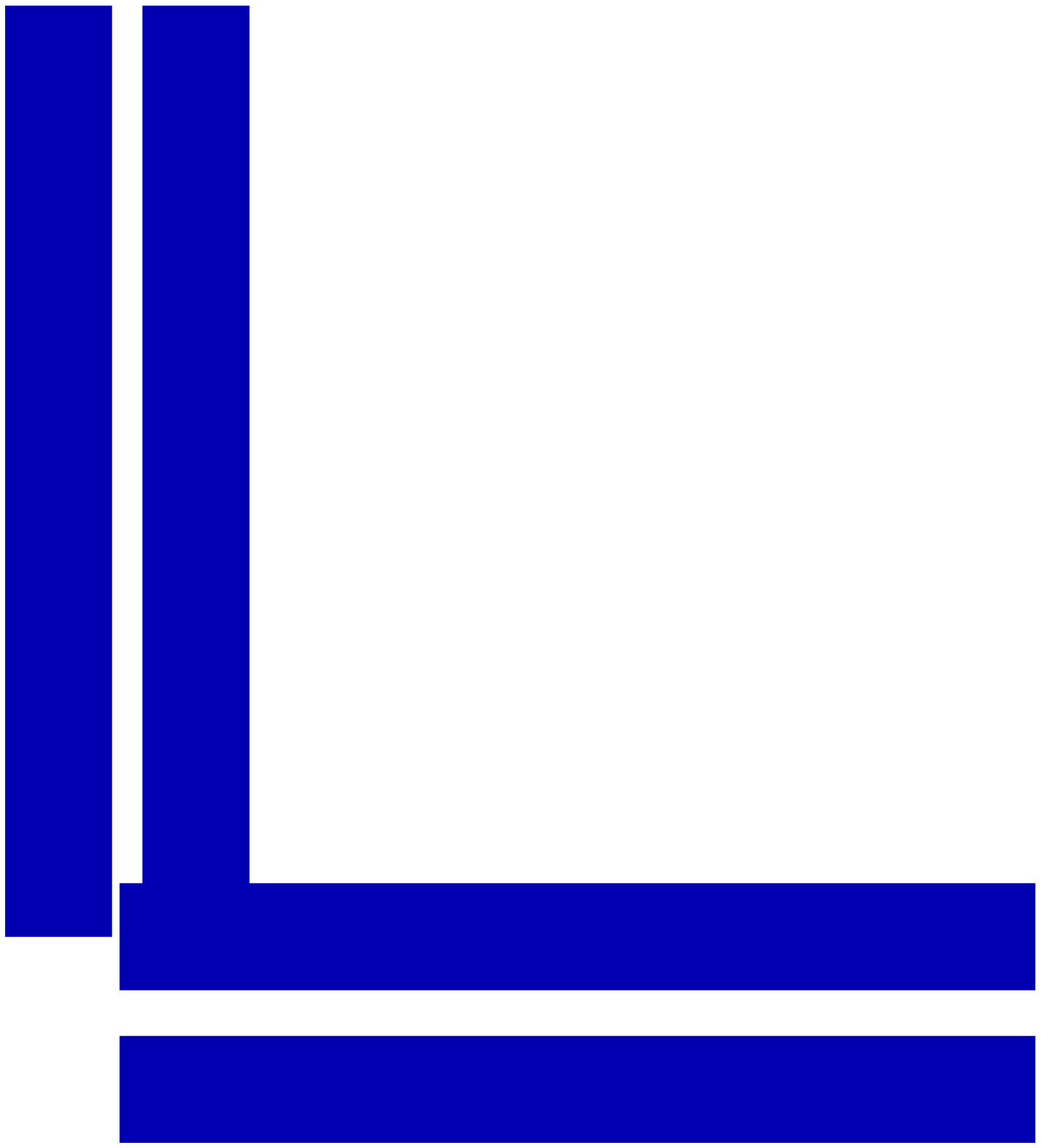}
\end{center}
\end{minipage}
&
$W^c_4$
&
&
&
\\
& & & & & \\
\hline
\end{tabular}
\caption{\label{tab2} Types of vertices in a fermion bag. The weights are given in Eqs.\ref{weights1}-\ref{weights7}.}
}

The simplest site is type-0 site where $n_x = 4$. Such a site forms its own bag since it is not connected to any dimers. It has a weight 
\begin{subequations}
\begin{equation}
\omega_B = W_0 = \kappa^{-4}.
\label{weights1}
\end{equation}
Next consider the type-1 site with $n_x = 3$ and $k_{x,s_1\alpha_1}=1$ where $s_1=\pm 1$ and $\alpha_1$ is one of four possible positive directions. The contribution to the weight of the fermion bag due to such a site is in the form of a Dirac tensor $(W_1)^{s_1\alpha_1}_{i_1;j_1}$ and is given by
\begin{eqnarray}
(W_1)^{s_1\alpha_1}_{i_1;j_1} &=& i \frac{\kappa^{-3}}{3!}
(S_{-s_1,\alpha_1})_{i_1 k_1} \varepsilon_{k_1 k_2 k_3 k_4} 
\varepsilon_{l_1 k_2 k_3 k_4} (S_{s_1,\alpha_1}^\dagger)_{l_1 j_1}
\nonumber \\
&=& 
i \kappa^{-3} \Big(S_{-s_1,\alpha_1}S^\dagger_{s_1,\alpha_1}\Big)_{i_1j_1}  = 0.
\end{eqnarray}
Hence a fermion bag cannot contain a type-1 vertex.

Next consider the type-2 vertex where $n_x=2$ and $k_{x,s_1\alpha_1} = k_{x,s_2\alpha_2}=1$. In this case the Dirac tensor associated with this site is of the form $(W_2)^{s_1\alpha_1,s_2\alpha_2}_{i_1 i_2;j_1 j_2}$ that contributes to $\omega_B$ is given by
\begin{eqnarray}
(W_2)^{s_1\alpha_1,s_2\alpha_2}_{i_1,i_2;j_1,j_2} &=& - \frac{\kappa^{-2}}{2!}
(S_{-s_1,\alpha_1})_{i_1 k_1} (S_{-s_2,\alpha_2})_{i_2 k_2} 
\varepsilon_{k_1 k_2 k_3 k_4} 
\varepsilon_{l_1 l_2 k_3 k_4} 
(S_{s_1,\alpha_1}^\dagger)_{l_1 j_1}(S_{s_2,\alpha_2}^\dagger)_{l_2 j_2}
\nonumber \\
&=& \kappa^{-2}
\Big(S_{-s_1,\alpha_1}S^\dagger_{s_2,\alpha_2}\Big)_{i_1j_2} 
\Big(S_{-s_2,\alpha_2}S^\dagger_{s_1,\alpha_1}\Big)_{i_2j_1} 
= \kappa^{-2}
(R^{s_1s_2}_{\alpha_1\alpha_2})_{i_1j_2}(R^{s_2s_1}_{\alpha_2\alpha_1})_{i_2j_1}
\nonumber \\
\end{eqnarray}
Note that if $\alpha_1 = \alpha_2$ and $s_1 = s_2$ the tensor is zero.

Next consider the type-3 site with $n_x=1$ and $k_{x,s_1\alpha_1} = k_{x,s_2\alpha_2} = k_{x,s_3\alpha_3} = 1$. Then
\begin{eqnarray}
(W_3)^{s_1\alpha_1,s_2\alpha_2,s_3\alpha_3}_{i_1,i_2,i_3;j_1,j_2,j_3} &=& -i \kappa^{-1}
(S_{-s_1,\alpha_1})_{i_1 k_1} (S_{-s_2,\alpha_2})_{i_2 k_2} (S_{-s_3,\alpha_3})_{i_3 k_3} 
\varepsilon_{k_1 k_2 k_3 k_4} 
\nonumber \\
&& \ \ \ \ \varepsilon_{l_1 l_2 l_3 k_4} 
(S_{s_1,\alpha_1}^\dagger)_{l_1 j_1}
(S_{s_2,\alpha_2}^\dagger)_{l_2 j_2}(S_{s_3,\alpha_3}^\dagger)_{l_3 j_3}
\nonumber \\
&=& - i \kappa^{-1} \Bigg\{
\Big(S_{-s_1,\alpha_1}S^\dagger_{s_2,\alpha_2}\Big)_{i_1j_2} 
\Big(S_{-s_2,\alpha_2}S^\dagger_{s_3,\alpha_3}\Big)_{i_2j_3} 
\Big(S_{-s_3,\alpha_3}S^\dagger_{s_1,\alpha_1}\Big)_{i_3j_1}
\nonumber \\ 
&&\ \ \ \ \ \ + 
\Big(S_{-s_1,\alpha_1}S^\dagger_{s_3,\alpha_3}\Big)_{i_1j_3}
\Big(S_{-s_3,\alpha_3}S^\dagger_{s_2,\alpha_2}\Big)_{i_3j_2} 
\Big(S_{-s_2,\alpha_2}S^\dagger_{s_1,\alpha_1}\Big)_{i_2j_1}\Bigg\}
\nonumber \\
&=& - i \kappa^{-1} \Big(
(R^{s_1s_2}_{\alpha_1\alpha_2})_{i_1j_2}(R^{s_2s_3}_{\alpha_2\alpha_3})_{i_2j_3}
(R^{s_3s_1}_{\alpha_3\alpha_1})_{i_3j_1}
\nonumber \\
&& \hskip1in +
(R^{s_1s_3}_{\alpha_1\alpha_3})_{i_1j_3}(R^{s_3s_2}_{\alpha_3\alpha_2})_{i_3j_2}
(R^{s_2s_1}_{\alpha_2\alpha_1})_{i_2j_1}
\Big)
\end{eqnarray}
Note that again all the dimers must be in different directions otherwise the site weight is zero.

Finally we can have a type-4 site with $n_x = 0$. In this case we have four directions given by $k_{x,s_1\alpha_1} = k_{x,s_2\alpha_2} = k_{x,s_3\alpha_3} = k_{x,s_4\alpha_4} = 1$. Now there are three possibilities: Type-4a is one in which all the four dimers are in different directions and then we get
\begin{eqnarray}
(W^a_4)^{s_1\alpha_1,s_2\alpha_2,s_3\alpha_3,s_4\alpha_4}_{i_1,i_2,i_3,i_4;j_1,j_2,j_3,j_4} &=& 
(S_{-s_1,\alpha_1})_{i_1 k_1} (S_{-s_2,\alpha_2})_{i_2 k_2} 
(S_{-s_3,\alpha_3})_{i_3 k_3} (S_{-s_4,\alpha_4})_{i_4 k_4} 
\varepsilon_{k_1 k_2 k_3 k_4} 
\nonumber \\
&& 
\ \ \ \ \varepsilon_{l_1 l_2 l_3 l_4} 
(S_{s_1,\alpha_1}^\dagger)_{l_1 j_1}
(S_{s_2,\alpha_2}^\dagger)_{l_2 j_2}
(S_{s_3,\alpha_3}^\dagger)_{l_3 j_3}
(S_{s_4,\alpha_4}^\dagger)_{l_4 j_4}
\nonumber \\
&=& \Big(
(R^{s_1s_2}_{\alpha_1\alpha_2})_{i_1j_2}
(R^{s_2s_1}_{\alpha_2\alpha_1})_{i_2j_1}
(R^{s_3s_4}_{\alpha_3\alpha_4})_{i_3j_4}
(R^{s_4s_3}_{\alpha_4\alpha_3})_{i_4j_3}
\nonumber \\
&&
- (R^{s_1s_2}_{\alpha_1\alpha_2})_{i_1j_2}
(R^{s_2s_3}_{\alpha_2\alpha_3})_{i_2j_3}
(R^{s_3s_4}_{\alpha_3\alpha_4})_{i_3j_4}
(R^{s_4s_1}_{\alpha_4\alpha_1})_{i_4j_1}
\nonumber \\
&&
-(R^{s_1s_2}_{\alpha_1\alpha_2})_{i_1j_2}
(R^{s_2s_4}_{\alpha_2\alpha_4})_{i_2j_4}
(R^{s_4s_3}_{\alpha_4\alpha_3})_{i_4j_3}
(R^{s_3s_1}_{\alpha_3\alpha_2})_{i_3j_1}
\nonumber \\
&&
+ (R^{s_1s_3}_{\alpha_1\alpha_3})_{i_1j_3}
(R^{s_3s_1}_{\alpha_3\alpha_1})_{i_3j_1}
(R^{s_2s_4}_{\alpha_2\alpha_4})_{i_2j_4}
(R^{s_4s_2}_{\alpha_4\alpha_2})_{i_4j_2}
\nonumber \\
&&
- (R^{s_1s_3}_{\alpha_1\alpha_3})_{i_1j_3}
(R^{s_3s_4}_{\alpha_3\alpha_4})_{i_3j_4}
(R^{s_4s_2}_{\alpha_4\alpha_2})_{i_4j_2}
(R^{s_2s_1}_{\alpha_2\alpha_1})_{i_2j_1}
\nonumber \\
&&
-(R^{s_1s_3}_{\alpha_1\alpha_3})_{i_1j_3}
(R^{s_3s_2}_{\alpha_3\alpha_2})_{i_3j_2}
(R^{s_2s_4}_{\alpha_2\alpha_4})_{i_2j_4}
(R^{s_4s_1}_{\alpha_4\alpha_1})_{i_4j_1}
\nonumber \\
&&
+ (R^{s_1s_4}_{\alpha_1\alpha_4})_{i_1j_4}
(R^{s_4s_1}_{\alpha_4\alpha_1})_{i_4j_1}
(R^{s_2s_3}_{\alpha_2\alpha_3})_{i_2j_3}
(R^{s_3s_2}_{\alpha_3\alpha_2})_{i_3j_2}
\nonumber \\
&&
- (R^{s_1s_4}_{\alpha_1\alpha_4})_{i_1j_4}
(R^{s_4s_3}_{\alpha_4\alpha_3})_{i_4j_3}
(R^{s_3s_2}_{\alpha_3\alpha_2})_{i_3j_2}
(R^{s_2s_1}_{\alpha_2\alpha_1})_{i_2j_1}
\nonumber \\
&&
-(R^{s_1s_4}_{\alpha_1\alpha_4})_{i_1j_4}
(R^{s_4s_2}_{\alpha_4\alpha_2})_{i_4j_2}
(R^{s_2s_3}_{\alpha_2\alpha_3})_{i_2j_3}
(R^{s_3s_1}_{\alpha_3\alpha_1})_{i_3j_1}
\Big)
\label{weights5}
\end{eqnarray}
Type-4b is the site where $k_{x,s_1\alpha_1} = 2$ and $k_{x,s_3\alpha_3} = k_{x,s_4\alpha_4}=1$. The above expression then simplifies to 
\begin{eqnarray}
(W^b_4)^{s_1\alpha_1,s_3\alpha_3,s_4\alpha_4}_{i_1,i_2,i_3,i_4;j_1,j_2,j_3,j_4} &=&
\frac{1}{2}\Big(
(R^{s_1s_3}_{\alpha_1\alpha_3})_{i_1j_3}
(R^{s_1s_4}_{\alpha_1\alpha_4})_{i_2j_4}
-
(R^{s_1s_3}_{\alpha_1\alpha_3})_{i_2j_3}
(R^{s_1s_4}_{\alpha_1\alpha_4})_{i_1j_4}
\Big)
\nonumber \\
&&
(R^{s_3s_1}_{\alpha_3\alpha_1})_{i_3j_1}
(R^{s_4s_1}_{\alpha_4\alpha_1})_{i_4j_2}
- 
(R^{s_3s_1}_{\alpha_3\alpha_1})_{i_3j_2}
(R^{s_4s_1}_{\alpha_4\alpha_1})_{i_4j_1}\Big)
\nonumber \\
&=& \frac{1}{2}(\tau_2)_{i_1 i_2}\ (\tau_2)_{j_1j_2}
[(R^{s_1s_3}_{\alpha_1\alpha_3})^T (\tau_2)
(R^{s_1s_4}_{\alpha_1\alpha_4})]_{j_3j_4}
[(R^{s_3s_1}_{\alpha_3\alpha_1}) (\tau_2)
(R^{s_4s_1}_{\alpha_4\alpha_1})^T]_{i_3i_4}.
\nonumber \\
\label{weights6}
\end{eqnarray}
Here the extra factor of $1/2$ is due to the fact that there are two dimers on one of the bonds and this leads to the an extra factor $1/(2!)^2$ present in Eq.(\ref{link}). This extra factor can be divided equally between the two vertices that the dimer connects. Type4-c site is obtained if $k_{x,s_1\alpha_1} = k_{x,s_3\alpha_3}=2$. In this case we get
\begin{eqnarray}
(W^c_4)^{s_1\alpha_1,s_3\alpha_3}_{i_1,i_2,i_3,i_4;j_1,j_2,j_3,j_4}
&=& \frac{1}{4} (F^{s_1 s_3}_{\alpha_1\alpha_3})^4(\tau_2)_{i_1 i_2}\ 
(\tau_2)_{j_1j_2}(\tau_2)_{i_3i_4}(\tau_2)_{j_3j_4}
\label{weights7}
\end{eqnarray}
\end{subequations}
Again the extra factor of $1/4$ is due to two double dimers and the factor $F^{s_1,s_3}_{\alpha_1,\alpha_3}$ comes from the $R$ terms. This completes the classification of all the vertices

\TABLE[t]{
\begin{tabular}{c|c|c|c}
\hline
Bag & $\omega_B$ & dimer representation & bag type\\
\hline
& & &\\
\begin{minipage}[c]{0.2\textwidth}
\begin{center}
\includegraphics[width=0.8\textwidth]{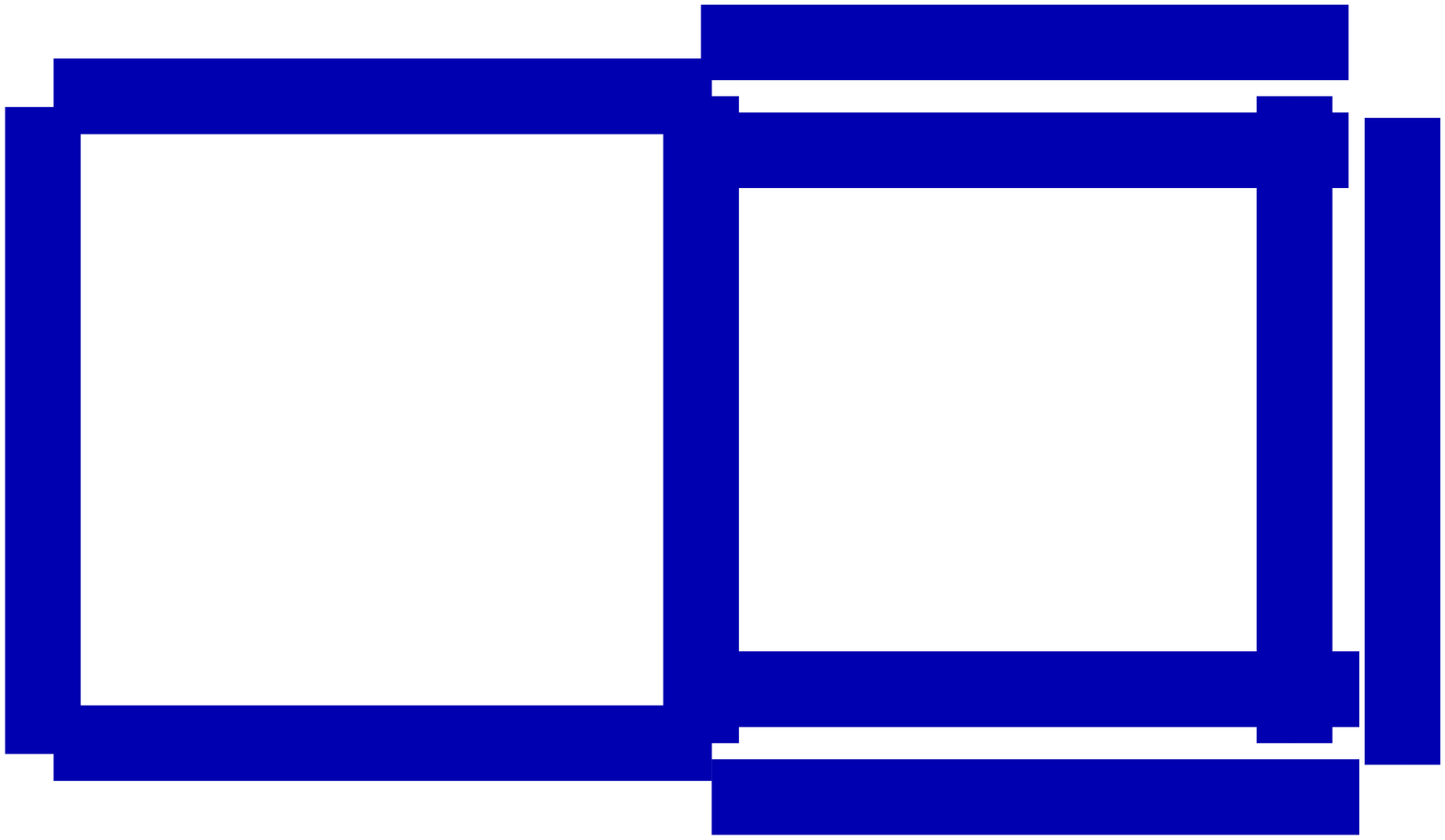}
\end{center}
\end{minipage}
&
$\frac{1}{64}\kappa^{-4}$ 
&
\begin{minipage}[c]{0.5\textwidth}
$\begin{array}{ccc}
(0,0,0,0;1,1) & (0,0,0,0;2,1) & (1,0,0,0;1,2) \cr 
(1,0,0,0;2,1) & (0,1,0,0;1,1) & (2,0,0,0;2,2) \cr
(1,1,0,0;1,2) & &
\end{array}$
\end{minipage}
&
$(0,0)$
\\
& & &\\
\hline
& & &\\
\begin{minipage}[c]{0.2\textwidth}
\begin{center}
\includegraphics[width=0.65\textwidth]{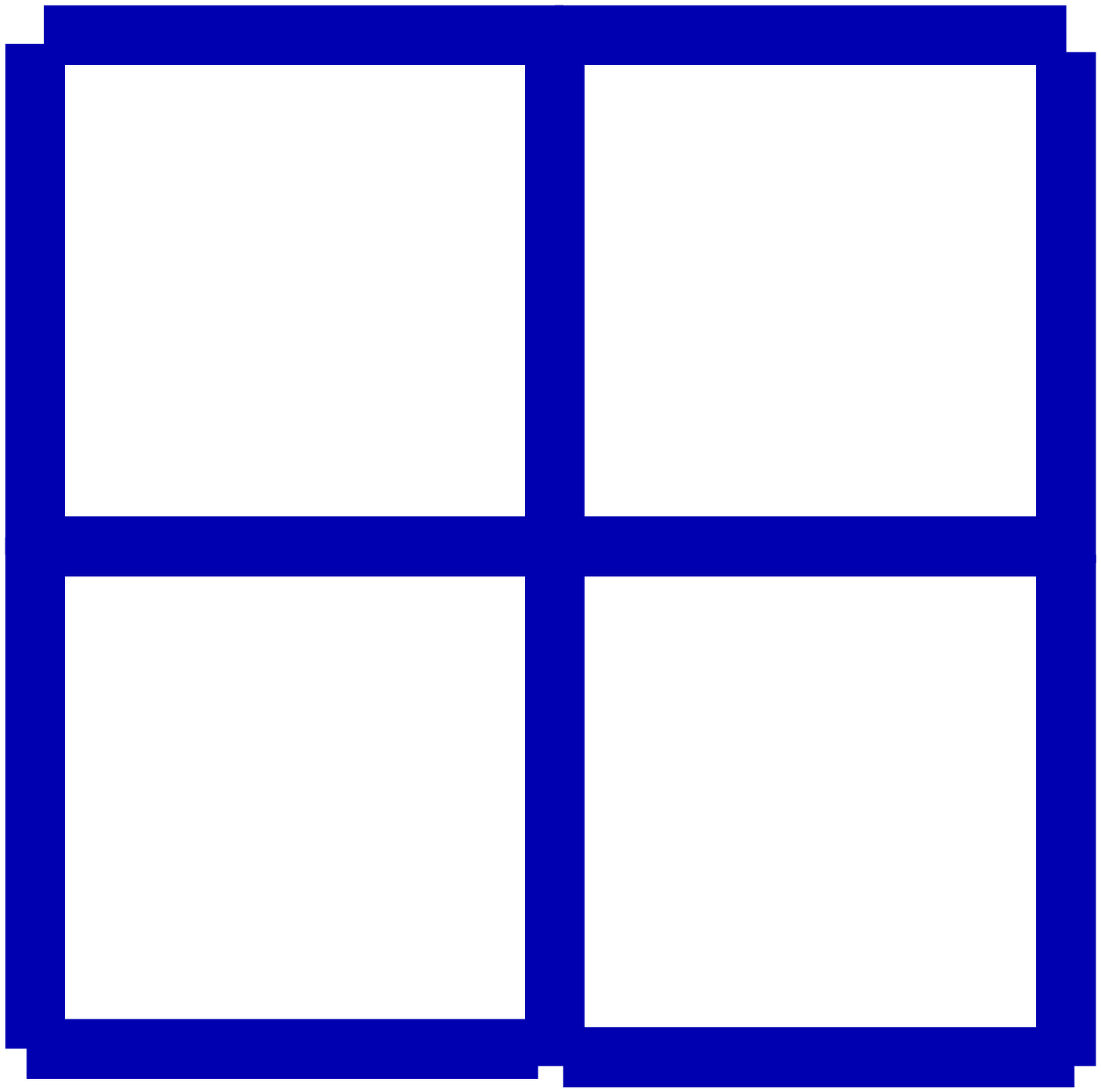}
\end{center}
\end{minipage}
&
$\frac{21}{128}\kappa^{-12}$ 
&
$\begin{array}{ccc}
(0,0,0,0;1,1) & (0,0,0,0;2,1) & (1,0,0,0;1,1) \cr
(1,0,0,0;2,1) & (0,1,0,0;1,1) & (0,1,0,0;2,1) \cr
(2,0,0,0;2,1) & (1,1,0,0;1,1) & (1,1,0,0;2,1) \cr
(0,2,0,0;1,1) & (2,1,0,0;2,1) & (1,2,0,0;1,1)
\end{array}$
&
$(4,1)$
\\
& & &\\
\hline
& & &\\
\begin{minipage}[c]{0.2\textwidth}
\begin{center}
\includegraphics[width=0.65\textwidth]{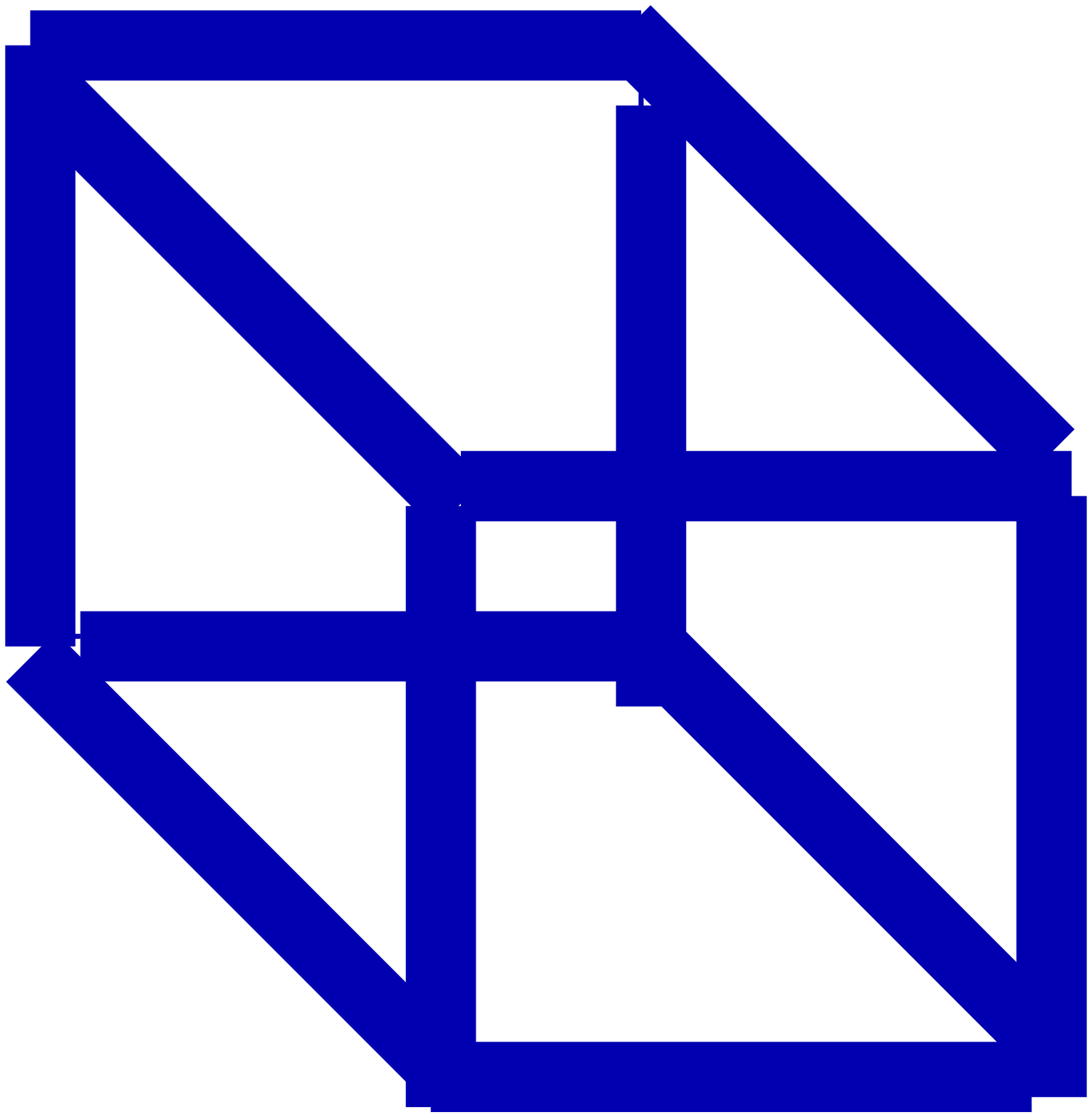}
\end{center}
\end{minipage}
&
$\frac{3}{32}\kappa^{-8}$ 
&
$\begin{array}{ccc}
(0,0,0,0;1,1) & (0,0,0,0;2,1) & (0,0,0,0;3,1) \cr
(1,0,0,0;2,1) & (1,0,0,0;3,1) & (0,1,0,0;1,1) \cr
(0,1,0,0;3,1) & (0,0,1,0;1,1) & (0,0,1,0;2,1) \cr
(1,1,0,0;3,1) & (1,0,1,0;2,1) & (0,1,1,0;1,1)
\end{array}$
&
$(8,0)$
\\
& & &\\
\hline
& & &\\
\begin{minipage}[c]{0.2\textwidth}
\begin{center}
\includegraphics[width=0.8\textwidth]{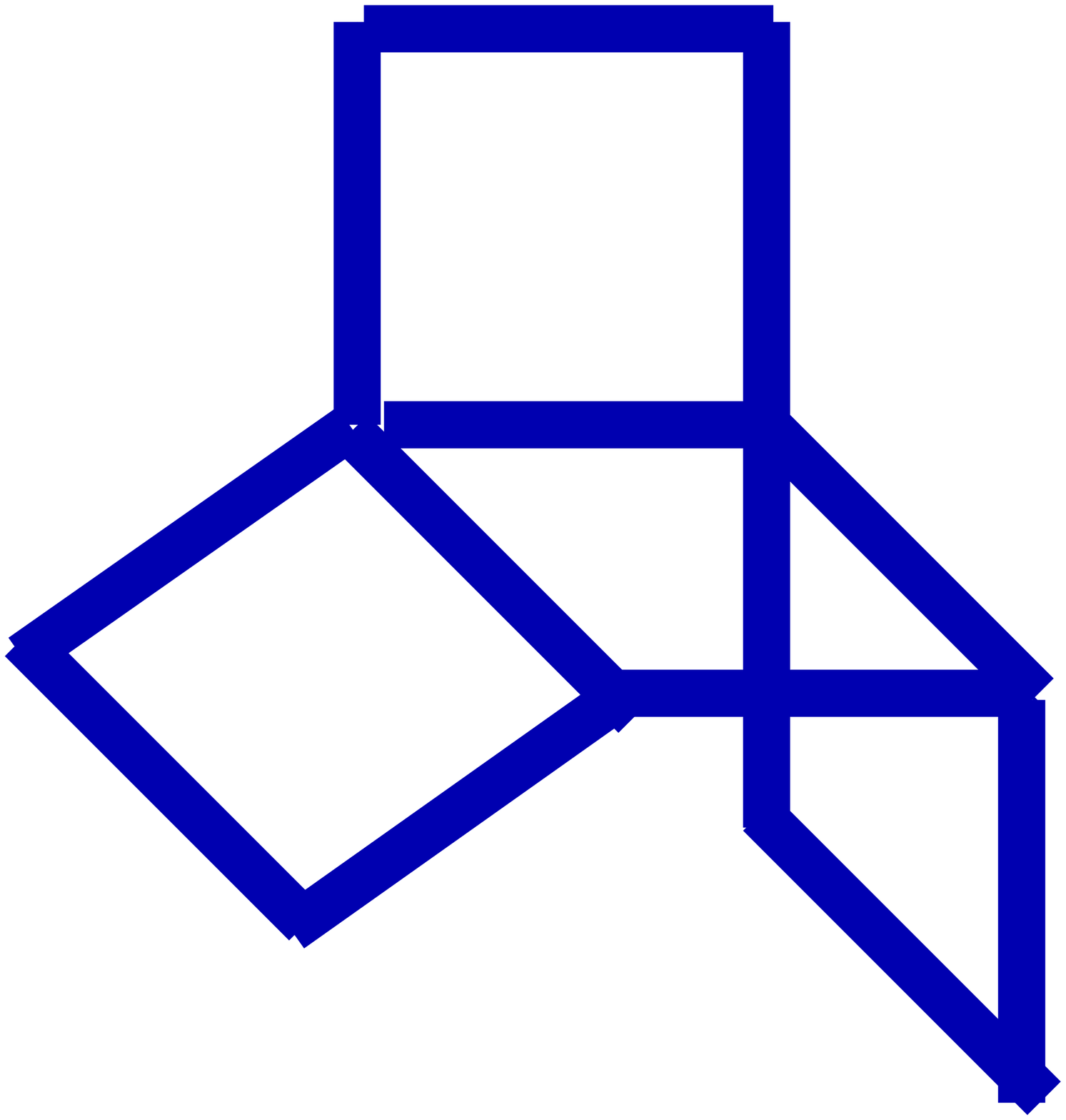}
\end{center}
\end{minipage}
&
$\frac{5}{2048}\kappa^{-14}$ 
&
$\begin{array}{ccc}
(0,0,0,0;1,1) & (0,0,0,0;2,1) & (0,0,0,0;4,1) \cr
(1,0,0,0;2,1) & (0,1,0,0;1,1) & (0,1,0,0;3,1) \cr
(0,1,0,0;4,1) & (0,0,0,1;2,1) & (1,0,3,0;2,1) \cr
(1,0,3,0;3,1) & (1,1,0,0;3,1) & (0,1,1,0;1,1) \cr
(1,1,3,0;3,1) & &
\end{array}$
&
$(2,2)$
\\
& & &\\
\hline
\end{tabular}
\caption{\label{bagexamples} Some small fermion bags and their weights.}
}

\TABLE[t]{
\begin{tabular}{c}
\hline
Bag 1 \\
\hline
$\begin{array}{cccccc}
(0,0,0,0;1,1) &
(0,0,0,0;2,1) &
(1,0,0,0;2,1) &
(1,0,0,0;3,2) &
(0,1,0,0;1,1) &
(1,0,1,0;4,1) \cr
(0,0,1,0;1,1) &
(0,0,1,0;4,1) &
(1,0,1,1;4,1) &
(0,0,1,1;2,1) &
(0,1,1,1;3,1) &
(0,0,1,2;1,1) \cr
(0,0,1,2;3,1) &
(0,0,2,1;2,1) &
(0,0,2,1;4,1) & & &\cr
\end{array}$
\\
\hline
Bag 2 \\
\hline
$\begin{array}{cccccc}
(0,0,0,0;2,1) &
(0,0,0,0;3,1) &
(0,1,0,0;2,1) &
(0,1,0,0;3,1) &
(0,0,1,0;1,1) &
(0,0,1,0;2,1)\cr
(0,2,0,0;3,1) &
(0,1,1,0;1,1) &
(0,1,1,0;2,1) &
(1,1,1,0;2,1) &
(1,0,0,0;2,1) &
(1,0,0,0;3,1)\cr
(1,1,0,0;4,1) &
(1,2,0,0;1,1) &
(1,2,0,0;3,1) &
(1,1,0,1;4,1) &
(2,2,0,0;3,1) &
(1,1,0,2;1,1)\cr
(1,0,0,2;1,1) &
(1,0,0,2;2,1) &
(2,1,1,0;2,1) &
(2,1,1,0;4,1) &
(2,0,0,2;2,1) &
(2,1,0,1;3,1)\cr
(2,1,0,1;4,1) & & & & &\cr
\end{array}$
\\
\hline
Bag 3 \\
\hline
$\begin{array}{cccccc}
(1,0,1,0;4,1) &
(0,0,1,0;1,1) &
(0,0,1,0;4,1) &
(1,0,1,1;4,1) &
(0,0,1,1;2,1) &
(0,1,1,1;3,1) \cr
(0,0,1,2;1,1) &
(0,0,1,2;3,1) &
(0,0,2,1;2,1) &
(0,0,2,1;4,1) & &\cr
\end{array}$
\\
\hline
\end{tabular}
\caption{\label{negbags} Examples of bags with negative weight (Bag 1, Bag 2) and zero weight (Bag 3). Bag 1 has a weight of $-1.220703125000 \times 10^{-4}$ and Bag 2 has a weight $-1.430511474609\times 10^{-6}$ at $\kappa = 1$.}
}

\section{Sign Problem with Fermion Bags}
\label{signfbag}

Using the rules of the previous section it is possible to compute the weights of fermion bags numerically. However, it is exponentially difficult to compute the weight when the bag contains many type-3 and type-4a sites. In order to make progress, we label the bags with the number of type-3 and type-4a vertices it contains. Thus, a bag of type $(n_3,n_4)$ contains $n_3$ type-3 sites and $n_4$ type-4a sites. The $(0,0)$ bags contain no type-3 and type-4a sites and will be referred to as {\em simple} bags. Bags in which either $n_3$ or $n_4$ is non-zero will be called {\em complex} bags. Below we will argue that this classification in terms of $(n_3,n_4)$ helps in understanding the origin of the sign problem.

By now it should be clear that every fermion bag can be uniquely represented through the dimers of the bag. We represent these dimers using the notation $(x_1,x_2,x_3,x_4;\alpha,k)$ where $x_i$ represent the four dimensional coordinates of the site inside the bag from which $k$ dimers emerge in the positive direction $\alpha$. Some examples of fermion bags, their dimer representation and their weights are given in table~\ref{bagexamples}. Although all the bags shown in the table have a positive weight we do find bags that have both zero weight and negative weights. However these bags are more complex. Two examples of negative weight bags and one example of zero weight bag are given in table~\ref{negbags} along with their weights : Bag-1 is a simple bag which contains twelve type-2 and two type-4b vertices. Bag-2 is a complex bag made up one type-4a, seventeen type-2, four type-3 vertices. Bag-3 is a simple loop bag with zero weight.

In order to understand the sign problem we have generated fermion bags of a fixed type at random on an $L^4$ lattice using a worm algorithm. In the case of simple bags, we exclude single site bags and plaquette bags for convenience. We then analyze the probability distribution of bags of a given type using the bag action density defined by
\begin{equation}
s_B = -\frac{1}{N_B} \log(|\omega_B|),
\end{equation}
where $N_B$ is the number of sites in the bag. In figure~\ref{fig9} we plot the distribution of $(0,0)$, $(2,0)$ and $(2,1)$ bags on the $2^4$ lattice with open boundary conditions as a function of $S_B$. For each type of bag we have generated $10^4$ bags. The left panel contains the distribution of positive weight bags while the right panel shows the distribution of the negative weight bags. We find that all simple bags (or $(0,0)$ type bags) turn out to have positive weights. On the other hand complex bags ($(2,0)$ and $(2,1)$ type bags) do contain negative weight bags. In the $(2,0)$ case we find $6987$ positive and $3013$ negative weight bags, while in the $(2,1)$ case we find $5983$ positive and $4017$ negative weight bags. We have repeated a similar analysis on a $5^4$ lattice where we have generated more that $3\times 10^4$ bags. These results are plotted in figure~\ref{fig10}. In this case a small number of simple bags do have negative weights. But the positive and negative weight complex bag distributions are almost identical for both $(2,0)$ and $(2,1)$ bags as can be seen from the figure.

Based on figures~\ref{fig9} and \ref{fig10} we conclude that as $n_3$ and $n_4$ increase (in other words as the bags become more complex) the distribution of positive and negative weight bags become more and more identical and hence the sign problem becomes severe. On the other hand simple bags are dominated by positive weight bags. Thus, we believe that to a very good approximation complex bags will cancel each other and the partition function is dominated by simple bags. Assuming this to be true an interesting effective model of strongly coupled QED emerges in which the partition function only contains simple bags. This model may share some of the physics of the original model. On the other hand it may be studied in its own right since it will have a much milder sign problem. We postpone this study to a future publication.

\FIGURE[h]{
\includegraphics[width=0.7\textwidth]{fig9.eps}
\caption{\label{fig9} Distribution positive weight bags (left panel) and negative weight bags (right panel) as a function of the action density $S_B$ on a $2^4$ hyper-cubic lattice with open boundary conditions. Three types of bags are shown: $(0,0)$-type (top) $(2,0)$-type (center) and $(2,1)$-type (bottom). See text for more details.}
}

\FIGURE[h]{
\includegraphics[width=0.7\textwidth]{fig10.eps}
\caption{\label{fig10} Distribution positive weight bags (left panel) and negative weight bags (right panel) as a function of the action density $S_B$ on a $5^4$ hyper-cubic lattice with open boundary conditions. Three types of bags are shown: $(0,0)$-type (top) $(2,0)$-type (center) and $(2,1)$-type (bottom). See text for more details.}
}

\section{Fermion Bags with Non-negative Weights}
\label{posfbags}

Can we construct a model of strongly coupled QED with Wilson fermions which is completely free of the sign problem in the fermion bag approach? In order to answer this question we identify fermion bags with non-negative weights. We find that there are three classes of fermion bags for which we can prove analytically that the Boltzmann weights are always non-negative. The first is the {\em trivial} bag consisting of a single site for which the weight is $W_0=\kappa^{-4}$. The second class are {\em loop} bags which have a loop topology. These bags only contain sites of type-2. Since they are closed loops of confined fermion and anti-fermion world lines, their weight is a square of a trace of an $SU(2)$ matrix and hence real and non-negative. Interestingly if we modify the original action to 
\begin{equation}
S = - \sum_{x,\alpha}\ 
\Big(
\psib_x\psi_x\ \psib_x \Gamma^\alpha_+ \mathrm{e}^{i\phi_{x,\alpha}}\psi_{x+\alpha} 
\ +\ 
\psib_{x+\alpha}\psi_{x+\alpha}\ 
\psib_{x+\alpha} \Gamma^\alpha_-  \mathrm{e}^{-i\phi_{x,\alpha}}\psi_{x}\Big) 
\ +\  \frac{1}{\kappa} \sum_x \psib_x\psi_x
\label{simplemodel}
\end{equation}
it is easy to argue that only trivial bags and loop bags are produced and the sign problem is completely solved. Since at small values of $\kappa$ type-3 and type-4 sites are naturally suppressed, this model may be a good approximation to the original model at small and intermediate values of $\kappa$. It most-likely contains the parity breaking phase transition of the original model \cite{Aoki:1983qi}. The hand-waving argument is as follows: At small values of $\kappa$ the loops are small while at large values loops proliferate the entire lattice and hence are naturally large. It is easy to show that on a finite lattice $\langle \psib \gamma_5 \psi\rangle = 0$ due to the parity symmetry. On the other hand the two point correlation function $\langle \psib_x \gamma_5 \psi_x \ \psib_y \gamma_5 \psi_y\rangle$ will be non-zero. This two point correlation function gets contribution through an open loop with the end points at $x$ and $y$. Intuitively, at small values of $\kappa$, since the loops will be small, the correlation function decays exponentially to zero for large separations. On the other hand at large values of $\kappa$, when the loops are large, the correlation function will decay as a power law and thus signaling the spontaneous breaking of parity. This phase transition can be studied efficiently using a worm-type algorithm. We postpone this study to the future. It would be interesting to understand the nature of this transition. Note that the above model will suffer from a severe sign problem in the conventional approach since additional auxiliary fields in addition to the usual gauge field will have to be introduced to convert the action into a fermion bi-linear. This is yet another example of a model which is solvable in the fermion-bag approach rather than the conventional approach.

The third class of bags with non-negative weights consist only of type-4 sites. These bags arise naturally when $\kappa=\infty$. The proof that the Boltzmann weight is non-negative is a bit more involved and relies on the bi-partite nature of the lattice. Let us briefly sketch the proof here. From Eq.(\ref{weights5}) we know that the contribution to the weight from each site within the bag comes from the tensor
\begin{eqnarray}
(W^a_4)^{s_1\alpha_1,s_2\alpha_2,s_3\alpha_3,s_4\alpha_4}_{i_1,i_2,i_3,i_4;j_1,j_2,j_3,j_4} &=& (S_{-s_1,\alpha_1})_{i_1 k_1} (S_{-s_2,\alpha_2})_{i_2 k_2} 
(S_{-s_3,\alpha_3})_{i_3 k_3} (S_{-s_4,\alpha_4})_{i_4 k_4} 
\varepsilon_{k_1 k_2 k_3 k_4} 
\nonumber \\
&& 
\ \ \ \ \Big[(S_{s_1,\alpha_1})_{j_1 l_1}
(S_{s_2,\alpha_2})_{j_2 l_2}
(S_{s_3,\alpha_3})_{j_3 l_3}
(S_{s_4,\alpha_4})_{j_4 l_4}\varepsilon_{l_1 l_2 l_3 l_4} \Big]^*
\end{eqnarray}
If we define
\begin{equation}
T^{s_1\alpha_1,s_2\alpha_2,s_3\alpha_3,s_4\alpha_4}_{i_1,i_2,i_3,i_4}
= (S_{-s_1,\alpha_1})_{i_1 k_1} (S_{-s_2,\alpha_2})_{i_2 k_2} 
(S_{-s_3,\alpha_3})_{i_3 k_3} (S_{-s_4,\alpha_4})_{i_4 k_4} 
\varepsilon_{k_1 k_2 k_3 k_4}
\end{equation}
we see that
\begin{equation}
W_4 = 
T^{s_1\alpha_1,s_2\alpha_2,s_3\alpha_3,s_4\alpha_4}_{i_1,i_2,i_3,i_4}
\
(T^{s_1\alpha_1,s_2\alpha_2,s_3\alpha_3,s_4\alpha_4}_{j_1,j_2,j_3,j_4})^*
\end{equation}
This structure of $W_4$ shows that, on a bi-partite lattice, the Boltzmann weight of the bag will be the square of the magnitude of a complex number obtained by tracing over the product of $T$'s on each site.

Although the above argument proves that all the fermion bags with type-4 vertices will have non-negative weights, as far as we know, a practical Monte Carlo algorithm seems impossible due to the fact that it will be exponentially difficult to compute the Boltzmann weight of large fermion bags. In a sense, the sign problem may still be hidden in this computational difficulty. 

\section{Conclusions}
\label{conc}

In this work we have constructed the fermion bag approach to strongly coupled lattice QED with one flavor of Wilson fermions in four dimensions. We found that at $\kappa=\infty$ all fermion bags have non-negative weights. On the other hand fermion bags with negative weights do exist and create a severe sign problem at intermediate values of $\kappa$. By classifying bags as simple and complex we could show that complex bags almost cancel each other in the partition function while simple bags are almost always positive and hence contribute to the partition function. This suggests a simple solution to the sign problem. We simply approximate the partition function as the sum of contributions form simple bags.

This approximate solution to the sign problem is similar in spirit to the meron cluster approach \cite{PhysRevLett.83.3116}. There special clusters called meron clusters appeared with equal weight but opposite sign in the partition function. Allowing meron clusters in the partition function would create a very severe sign problem. However, since they come with exactly equal weight and opposite signs, they cancel exactly and thus the sign problem was solved completely. In the current situation, the cancellation of complex bags is only approximate and suggestive. So, while we cannot justify rigorously that it is correct to ignore them in the partition function we believe it to be correct. In the future it would be interesting to study the partition function generated by simple bags alone.

Finally, we have also constructed a simpler model (Eq.~(\ref{simplemodel})) that consists of loop bag and does not suffer from the sign problem. Simple arguments suggest that this model contains two phases : a parity symmetric phase at small values of $\kappa$ and a phase where parity is spontaneously broken at large values of $\kappa$. It would be interesting to study the nature of this phase transition in three dimensions.

\acknowledgments

We would like to thank Urs Wenger for discussions about the solution to the sign problem with Wilson fermions in two and three dimensions. This work was supported in part by the Department of Energy grant DE-FG02-05ER41368. S.C. wishes to acknowledge the Aspen Center for Physics for support where part of this work was accomplished. A. Li would like to thank Ming Gong for useful discussions.

\appendix

\section{Fermion Determinant versus Fermion Bags}

In this appendix we provide checks that confirm the correctness of the rules that we constructed in section \ref{fbagrules}, to compute the weights of fermion bags. We compute the partition function on a small lattice by integrating $\mathrm{Det}(D_w)$ over the gauge fields exactly and identifying contributions from each of the fermion bags that are produced in the process. For simplicity we choose a $3 \times 2 $ and a $3\times 3$ lattice on the $xy-plane$ as shown in figure~\ref{fig:index_label}. The fermion bags that are produced in these small lattices already capture all types of vertices of table~\ref{tab2}. We label each lattice site by an index $i$ and assign different values $\kappa=\kappa_i$ to each site. The partition function will then be polynomial in $\kappa_i$'s such that powers of $\kappa_i$ are related to the number of monomers on site $i$. This helps identify terms in the partition function as weights of the fermion bags. Finally we set $\kappa_0 = \kappa_1 = \kappa_2= \cdots = \kappa$ to compute the weight of the bag.
\FIGURE[h]{
\begin{tabular}{cc}
\begin{minipage}[c]{0.4\textwidth}
\includegraphics[scale=0.3]{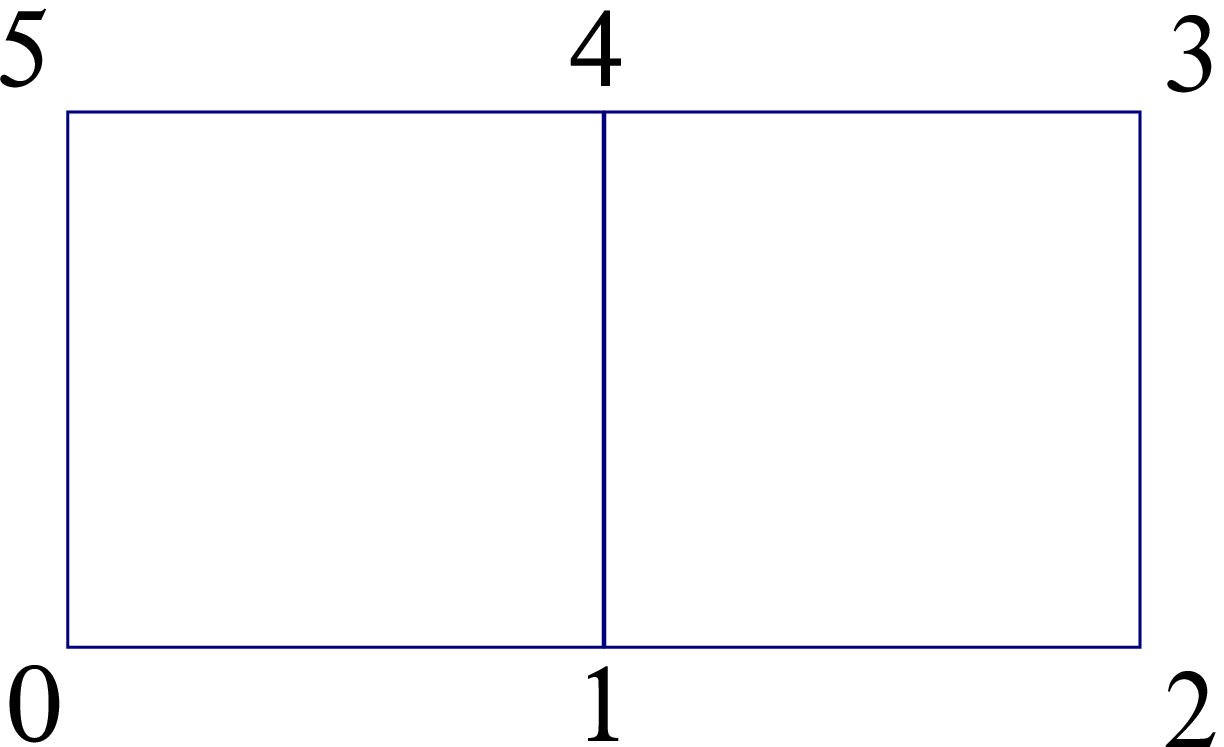} 
\end{minipage}
&
\begin{minipage}[c]{0.4\textwidth}
\includegraphics[scale=0.3]{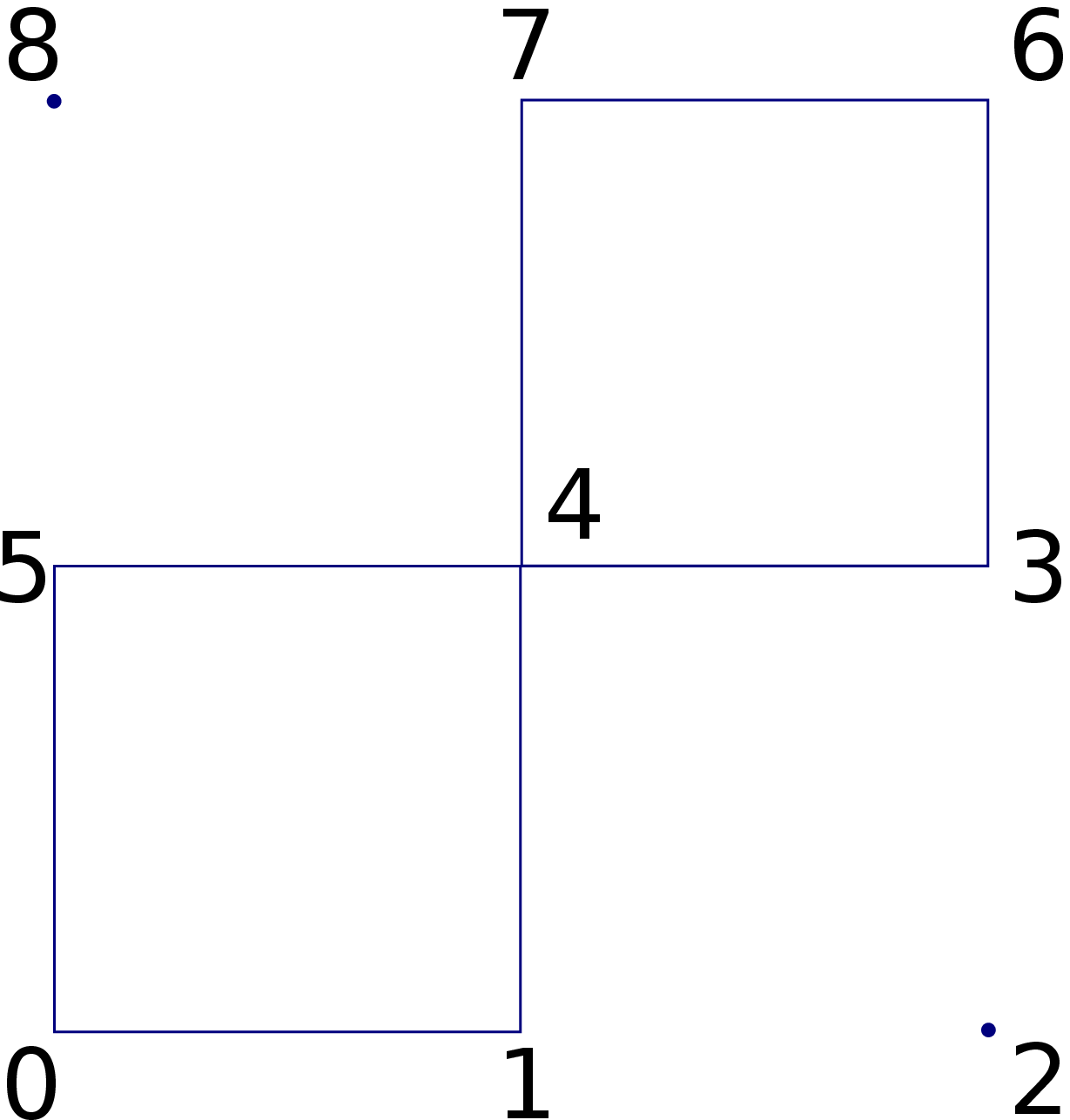}
\end{minipage}
\end{tabular}
\caption{Lattices on which we compute the partition function exactly. The site labels are used to label different values of $\kappa$ which help to identify the contribution to a particular fermion bag.
\label{fig:index_label}}
}

On a $3\times 2$ lattice we get find the partition function to be a sum of 11 terms given by
\begin{eqnarray}
Z & = & \frac{1}{256} + \frac{1}{256 \kappa _2^4 \kappa _3^4}+\frac{1}{256 \kappa _0^4 \kappa _5^4} + \frac{1}{64 \kappa _2^2 \kappa _3^2}+\frac{1}{64 \kappa _0^2 \kappa _5^2} \nonumber + \frac{1}{64 \kappa _0^2 \kappa _2^2 \kappa _3^2 \kappa_5^2} \nonumber \\
  & + & \frac{3}{16 \kappa _0^2 \kappa _1 \kappa _2^2 \kappa _3^2 \kappa _5^2 \kappa _4} + \frac{1}{4 \kappa _0^2 \kappa _1^2 \kappa _2^2 \kappa _3^2 \kappa _5^2 \kappa _4^2} + \frac{1}{4 \kappa _0^2 \kappa_1^2 \kappa _2^4 \kappa _3^4 \kappa _5^2 \kappa _4^2} + \frac{1}{4 \kappa _0^4 \kappa _1^2 \kappa _2^2 \kappa _3^2 \kappa _5^4 \kappa_4^2} \nonumber \\
  & + &\frac{1}{\kappa _0^4 \kappa _1^4 \kappa _2^4 \kappa _3^4 \kappa _5^4 \kappa _4^4}
\label{eq:31bags}
\end{eqnarray}
Collecting the terms into 8 categories each of which contributes to the weight of a fermion bag, we can compute these bag weights. In the top eight rows of table~\ref{tb:bags} we compare the weights computed by this method with the one computed using the fermion bag rules of section \ref{fbagrules}. In this calculation we find all vertices listed in table~\ref{tab2} except for type-4a vertex. To find a bag with type-4a vertex we compute the partition function on a $3 \times 3$ lattice. In order to simplify the calculation, we set the links between sites 2 and 1, 2 and 3, 8 and 5, 8 and 7 to be zero. Then the partition function contains 6 terms and are given by 
\begin{eqnarray}
Z & = & \frac{1}{256 \kappa _2^4 \kappa _3^4 \kappa _6^4 \kappa _7^4 \kappa _8^4}+\frac{3}{64 \kappa _0^2 \kappa _1^2 \kappa _2^4 \kappa _3^2 \kappa
   _6^2 \kappa _7^2 \kappa _8^4 \kappa _5^2}+\frac{1}{4 \kappa _0^2 \kappa _1^2 \kappa _2^4 \kappa _3^4 \kappa _4^2 \kappa _6^4 \kappa _7^4
   \kappa _8^4 \kappa _5^2} \nonumber \\
   & + & \frac{1}{4 \kappa _0^4 \kappa _1^4 \kappa _2^4 \kappa _3^2 \kappa _4^2 \kappa _6^2 \kappa _7^2 \kappa _8^4 \kappa
   _5^4}+\frac{1}{256 \kappa _0^4 \kappa _1^4 \kappa _2^4 \kappa _8^4 \kappa _5^4}+\frac{1}{\kappa _0^4 \kappa _1^4 \kappa _2^4 \kappa _3^4
   \kappa _4^4 \kappa _6^4 \kappa _7^4 \kappa _8^4 \kappa _5^4}
\label{eq:33bags}
\end{eqnarray}
These can be divided into 4 categories, 3 of them contribute to bags that have already been enumerated within the top eight rows of table~\ref{tb:bags}. The only bag which includes type-4a vertex is given in the last row of table~\ref{tb:bags}. These results confirm the rules constructed in section \ref{fbagrules}.
\TABLE[h]{
\begin{tabular}{c||c|c}
\hline  
Bag Diagram & Weight & Weight \\
& Determinant Approach  & Bag Approach \\
\hline
& & \\
\begin{minipage}[c]{0.4\textwidth}
\begin{center}
\includegraphics[width=0.3\textwidth]{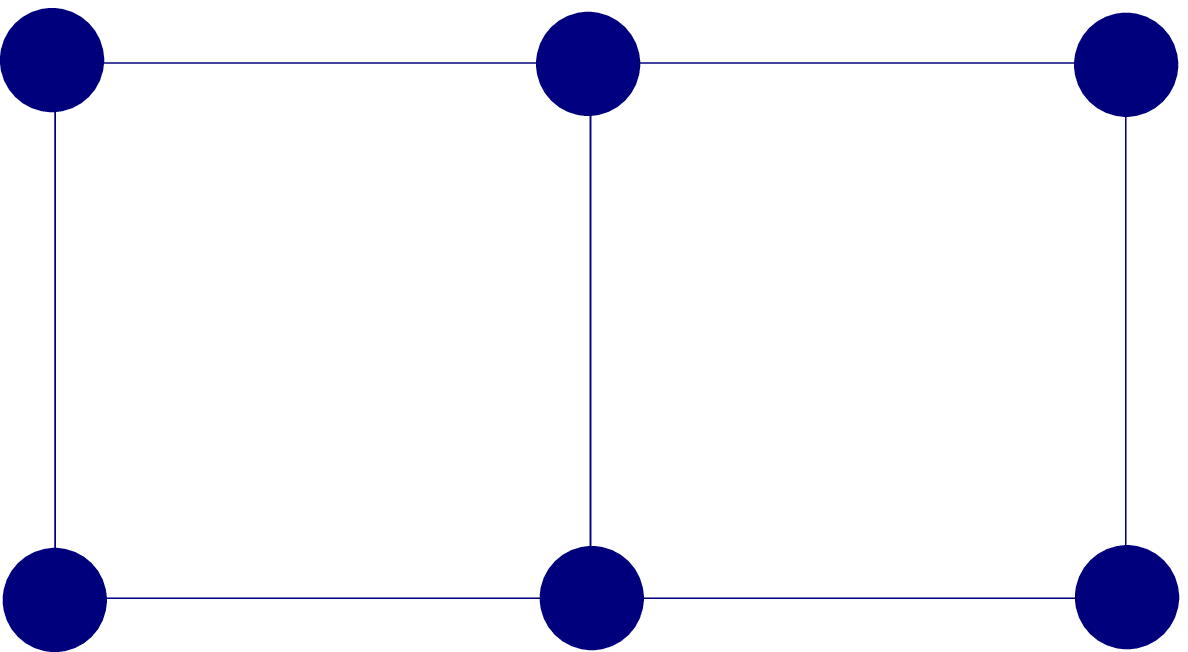}
\end{center}
\end{minipage} 
& $\displaystyle \frac{1}{\kappa _0^4 \kappa _1^4 \kappa _2^4 \kappa _3^4 \kappa _5^4 \kappa _4^4}$  & $\displaystyle \frac{1}{\kappa^{24}}$ \\
&  & \\ 
\hline  
& & \\
\begin{minipage}[c]{0.4\textwidth}
\begin{center}
\includegraphics[width=0.3\textwidth]{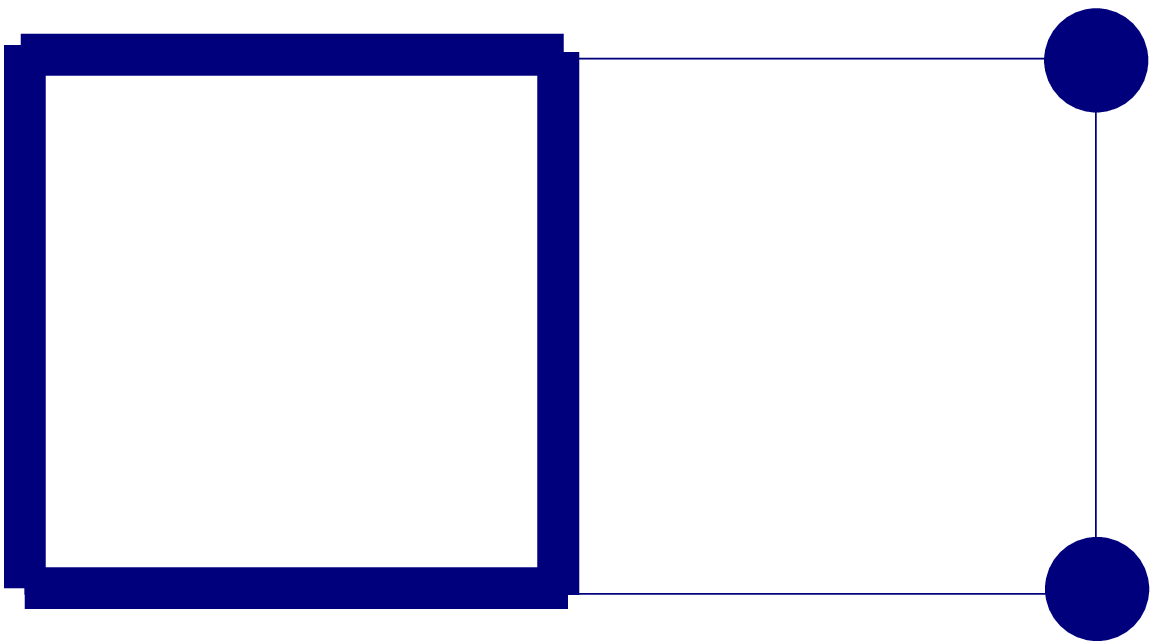}
\hskip0.2in
\includegraphics[width=0.3\textwidth]{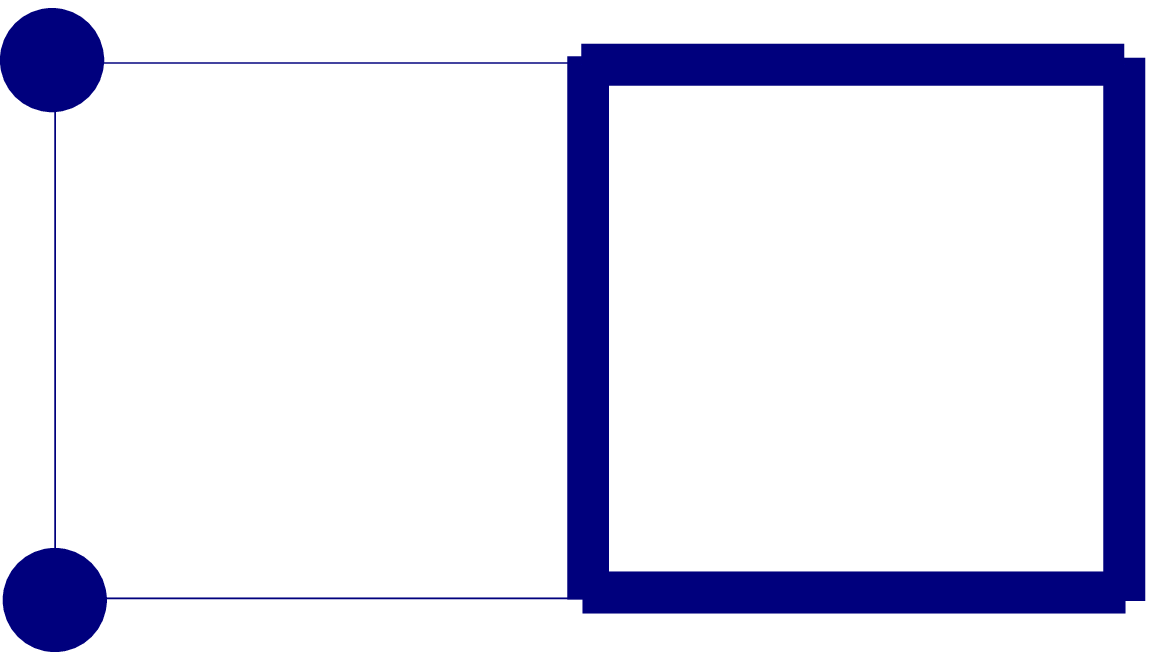}
\end{center}
\end{minipage} 
& $\displaystyle \frac{1}{4 \kappa _0^2 \kappa_1^2 \kappa _2^4 \kappa _3^4 \kappa _5^2 \kappa _4^2}, \frac{1}{4 \kappa _0^4 \kappa _1^2 \kappa _2^2 \kappa _3^2 \kappa _5^4 \kappa_4^2}$  & $\displaystyle \frac{1}{4\kappa^{16}}$ \\
&  & \\
\hline 
&  & \\
\begin{minipage}[c]{0.4\textwidth}
\begin{center}
\includegraphics[width=0.3\textwidth]{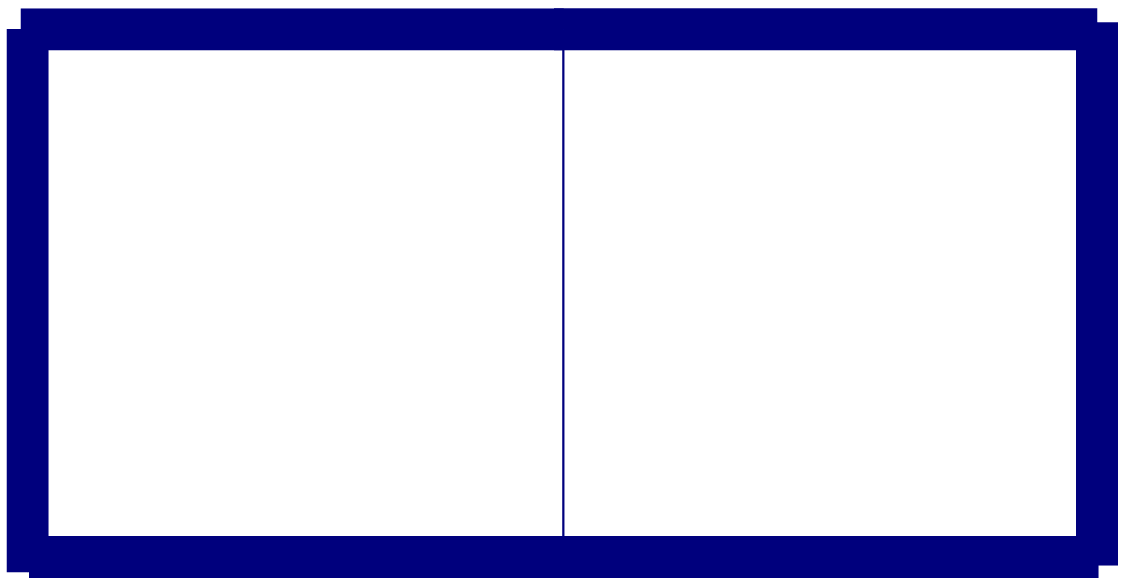}
\end{center}
\end{minipage} 
& $\displaystyle \frac{1}{4 \kappa _0^2 \kappa _1^2 \kappa _2^2 \kappa _3^2 \kappa _5^2 \kappa _4^2} $ & $\displaystyle \frac{1}{4\kappa^{12}}$\\
& & \\
\hline
& & \\
\begin{minipage}[c]{0.4\textwidth}
\begin{center}
\includegraphics[width=0.3\textwidth]{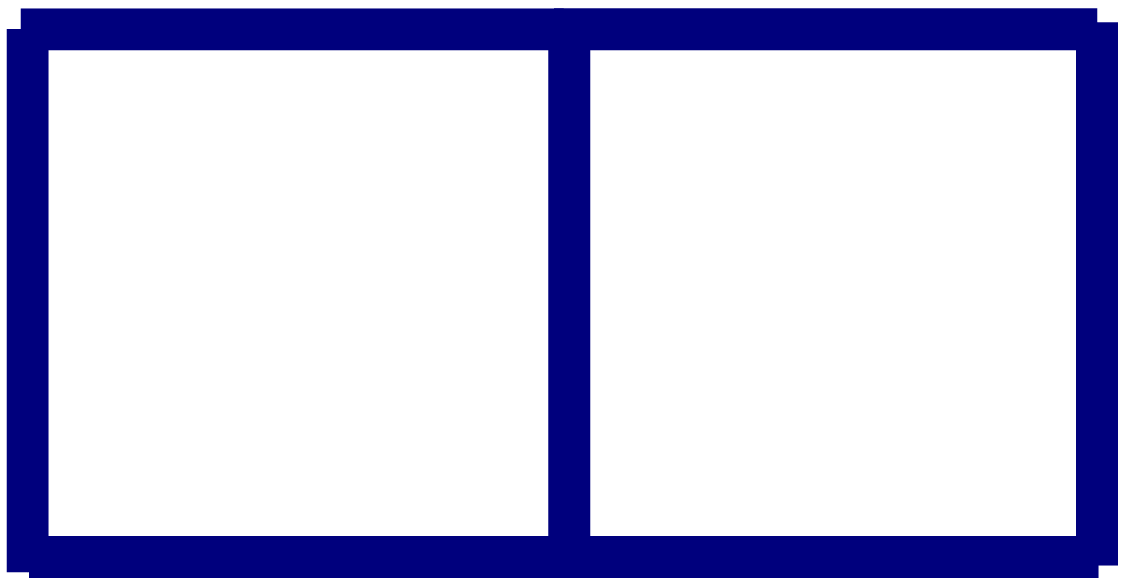}
\end{center}
\end{minipage}  
& $\displaystyle \frac{3}{16 \kappa _0^2 \kappa _1 \kappa _2^2 \kappa _3^2 \kappa _5^2 \kappa _4} $ & $\displaystyle \frac{3}{16\kappa^{10}}$ \\
& & \\
\hline 
& & \\
\begin{minipage}[c]{0.4\textwidth}
\begin{center}
\includegraphics[width=0.3\textwidth]{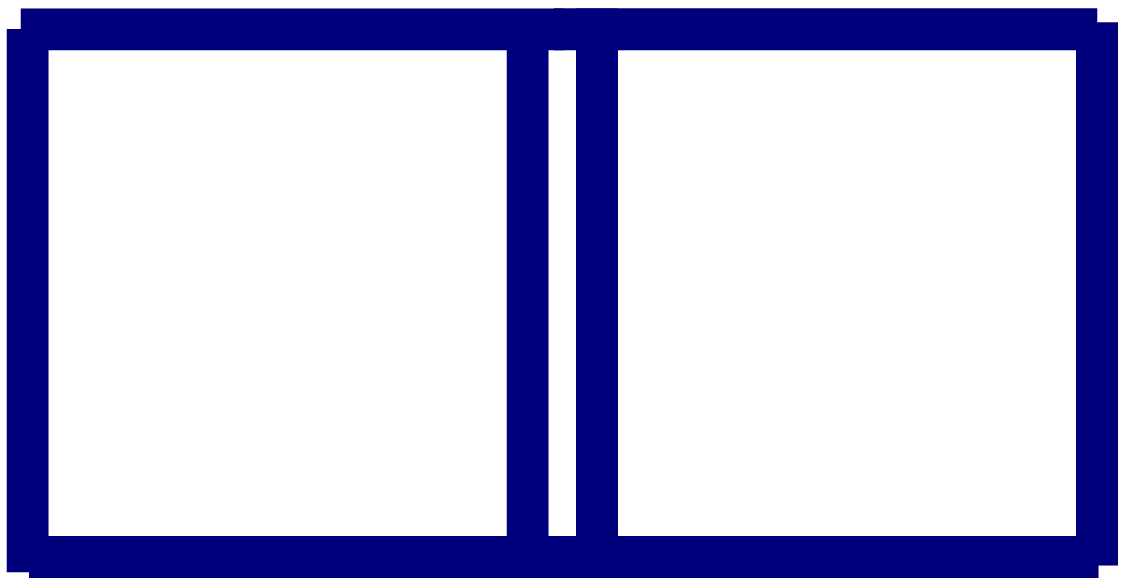}
\end{center}
\end{minipage}  
& $\displaystyle \frac{1}{64 \kappa _0^2 \kappa _2^2 \kappa _3^2 \kappa_5^2} $ & $\displaystyle \frac{1}{64\kappa^{8}}$ \\
& &\\
\hline  
& & \\
\begin{minipage}[c]{0.4\textwidth}
\begin{center}
\includegraphics[width=0.3\textwidth]{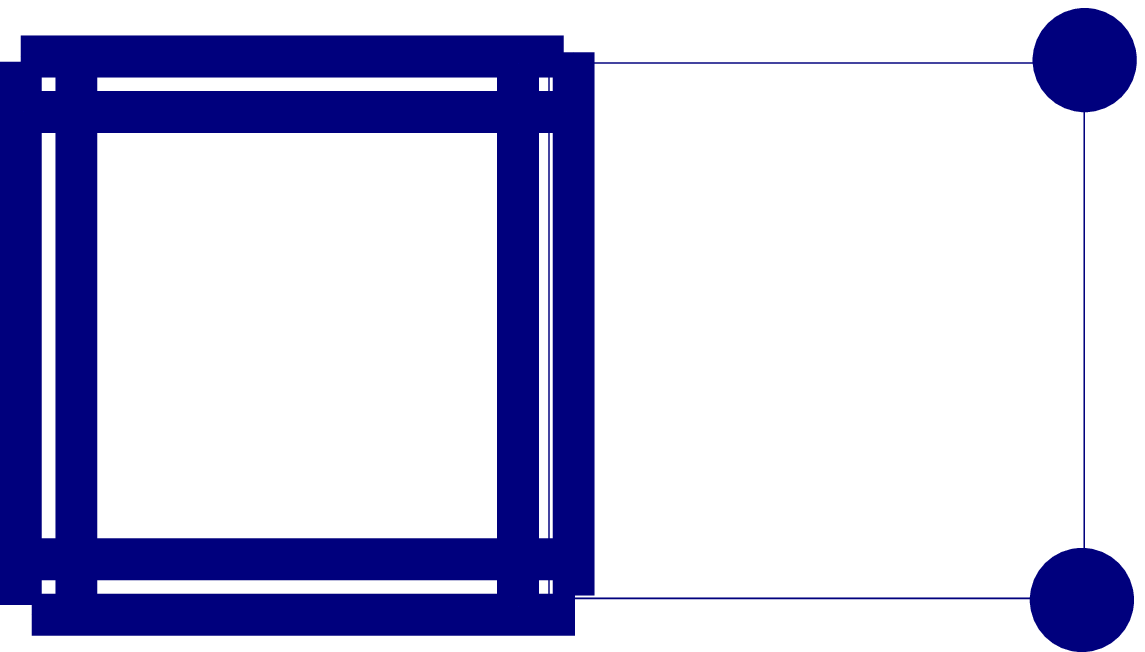}
\hskip0.2in
\includegraphics[width=0.3\textwidth]{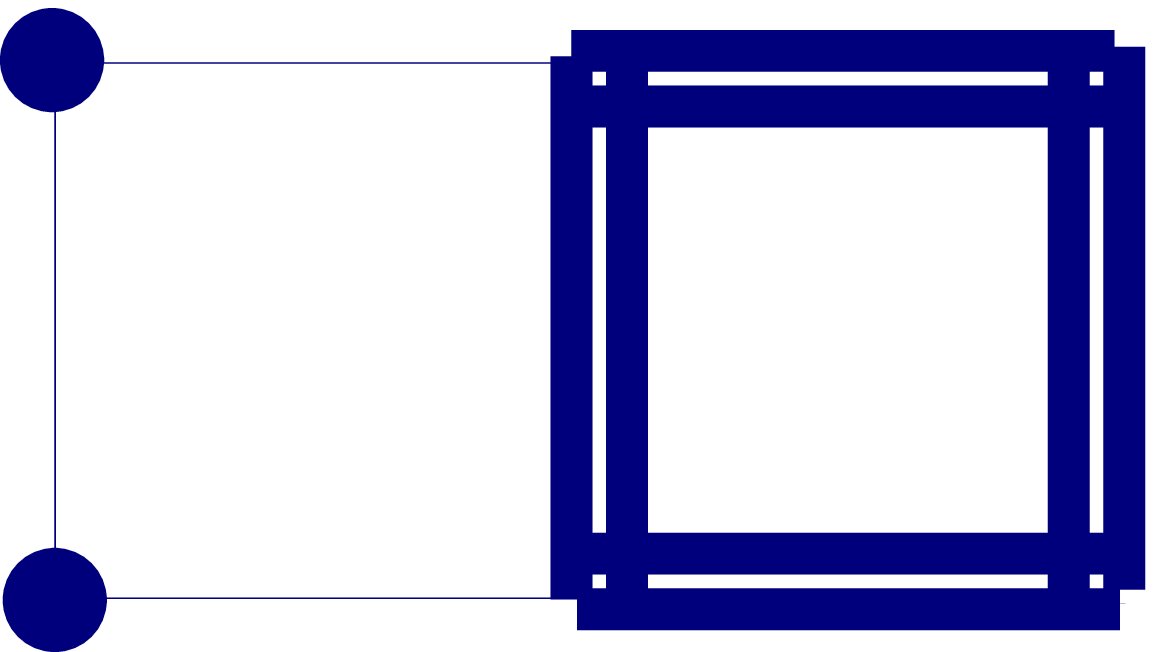}
\end{center}
\end{minipage}  
& $\displaystyle \frac{1}{256 \kappa _2^4 \kappa _3^4},\frac{1}{256 \kappa _0^4 \kappa _5^4} $  & $\displaystyle \frac{1}{256\kappa^{8}}$ \\
&  & \\
\hline 
&  &  \\
\begin{minipage}[c]{0.4\textwidth}
\begin{center}
\includegraphics[width=0.3\textwidth]{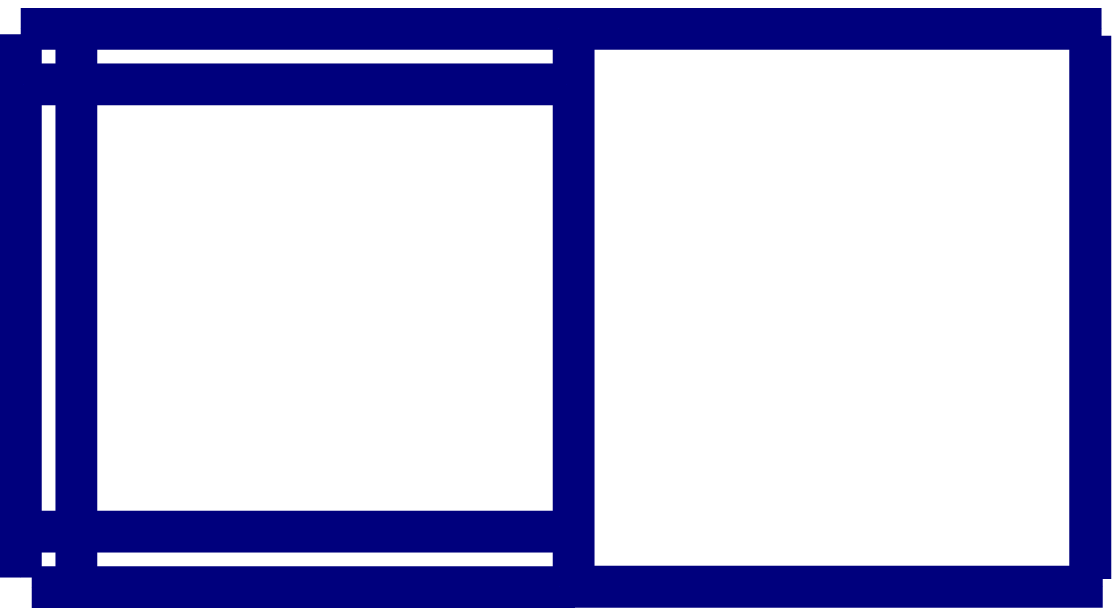}
\hskip0.2in
\includegraphics[width=0.3\textwidth]{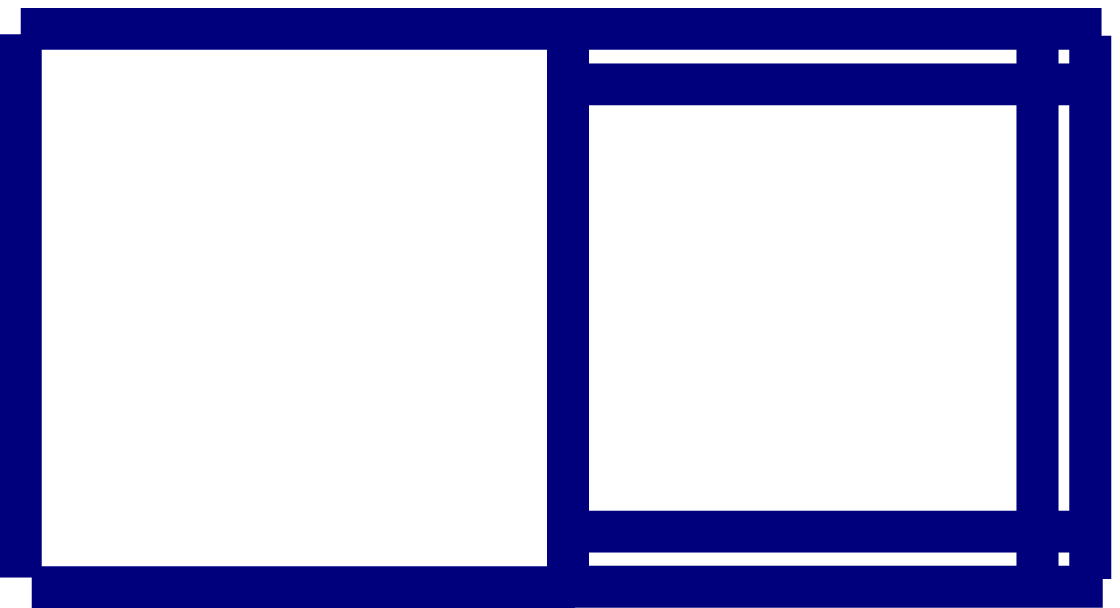}
\end{center}
\end{minipage}  
& $\displaystyle \frac{1}{64 \kappa _2^2 \kappa _3^2},\frac{1}{64 \kappa _0^2 \kappa _5^2} $  & $\displaystyle \frac{1}{64\kappa^{4}}$ \\
&  & \\
\hline  
&  &  \\
\begin{minipage}[c]{0.4\textwidth}
\begin{center}
\includegraphics[width=0.3\textwidth]{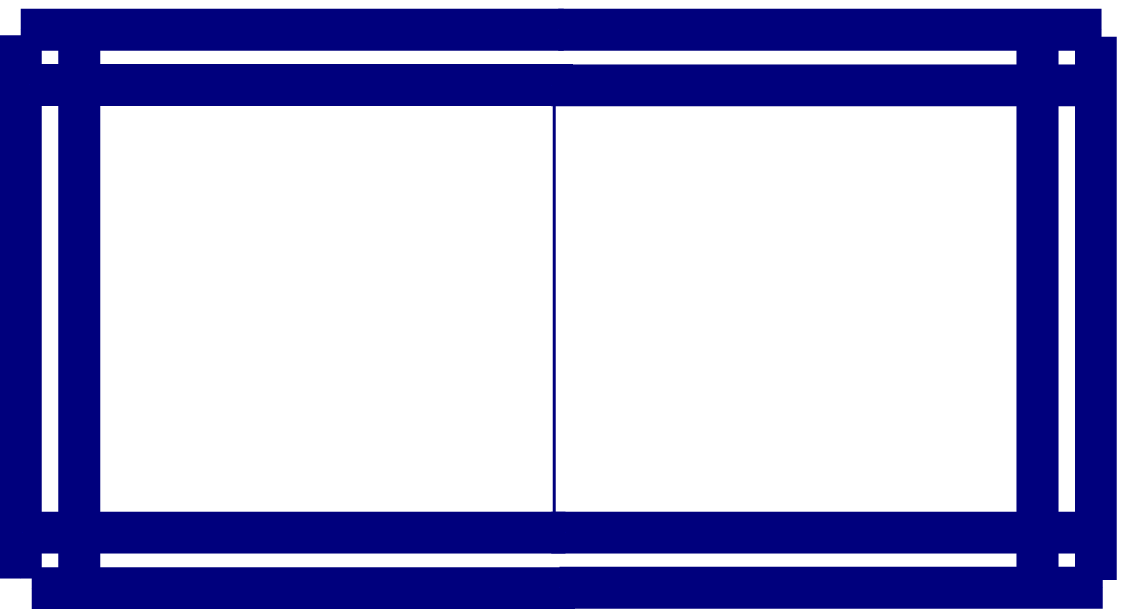}
\end{center}
\end{minipage}  
& $\displaystyle \frac{1}{256}$  & $\displaystyle \frac{1}{256}$ \\
&  & \\
\hline
&  &  \\
\begin{minipage}[c]{0.4\textwidth}
\begin{center}
\includegraphics[width=0.3\textwidth]{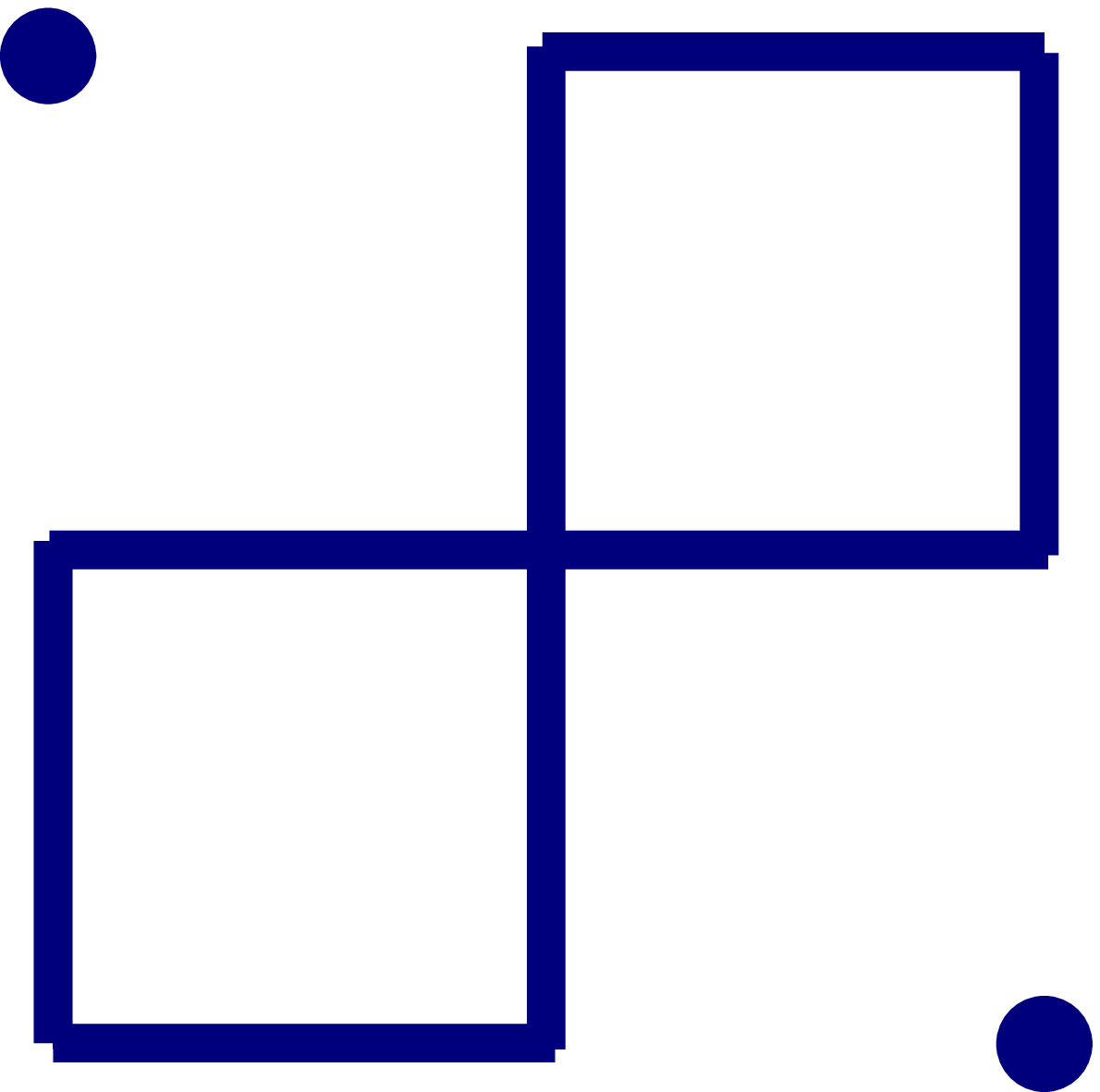}
\end{center}
\end{minipage}  
& $\displaystyle \frac{3}{64 \kappa _0^2 \kappa _1^2 \kappa _2^4 \kappa _3^2 \kappa_6^2 \kappa _7^2 \kappa _8^4 \kappa _5^2}$  & $\displaystyle \frac{3}{64\kappa^{20}}$  \\
&  & \\ 
\hline 
\end{tabular}
\caption{Comparison between the determinant approach and the bag approach on a $3\times2$ lattice (top 8 rows) and on a $3\times 3$ lattice (bottom row) with open boundary conditions. Each diagram corresponds a unique term in the partition function.\label{tb:bags}}
}

\clearpage

\bibliography{ref}

\bibliographystyle{JHEP}

\end{document}